\documentclass[useAMS,usenatbib,usegraphicx]{mn2e}

\title[A Detailed Study of 2S 0114+650 with the \emph{RXTE}]{A Detailed Study of 2S 0114+650 with the \emph{RXTE}}
\author[S. A. Farrell et al.]{S. A. Farrell$^{1,2}$\thanks{E-mail:
sean.farrell@cesr.fr}, R. K. Sood$^{2}$, P. M.
O'Neill$^{3,4}$ and S. Dieters$^{5}$\\
$^{1}$Centre d'\'{E}tude Spatiale des Rayonnements, CNRS/UPS, 9
Avenue du Colonel Roche, 31028 Toulouse Cedex 04, France\\
$^{2}$School of PEMS, UNSW@ADFA, Canberra ACT 2600, Australia\\
$^{3}$Astrophysics Group, Imperial College, Prince Consort Road, London SW7 2AZ, UK\\
$^{4}$School of Computing \& Mathematics, Charles Sturt
University, Locked Bag 588, Wagga Wagga NSW 2678, Australia\\
$^{5}$School of Mathematics and Physics, University of Tasmania,
Private Bag 21, Hobart Tasmania 7001, Australia}
\begin{document}

\date{Released 2007 Xxxxx XX}

\pagerange{\pageref{firstpage}--\pageref{lastpage}} \pubyear{2002}

\maketitle

\label{firstpage}

\begin{abstract}
We present the results of a detailed study of the high mass X-ray
binary 2S 0114+650 made with the pointed instruments onboard the
\emph{Rossi X-ray Timing Explorer}. The spectral and temporal
behaviour of this source was examined over the pulse, orbital, and
super-orbital timescales, covering $\sim$2 cycles of the 30.7 d
super-orbital modulation. Marginal evidence for variability of the
power law photon index over the pulse period was identified,
similar to that observed from other X-ray pulsars. If this
variability is real it can be attributed to a varying viewing
geometry of the accretion region with the spin of the neutron
star. Variability of the neutral hydrogen column density over the
orbital period was observed, which we attribute to the line of
sight motion of the neutron star through the dense circumstellar
environment. A reduction in the power law photon index was
observed during the orbital maximum, which we speculate is due to
absorption effects as the neutron star passes behind a heavily
absorbing region near the base of the supergiant companion's wind.
No significant variability of the column density was observed over
the super-orbital period, indicating that variable obscuration by
a precessing warp in an accretion disc is not the mechanism behind
the super-orbital modulation. In contrast, a significant increase
in the power law photon index was observed during the
super-orbital minimum. We conclude that the observed super-orbital
modulation is tied to variability in the mass accretion rate due
to some as yet unidentified mechanism.

\end{abstract}

\begin{keywords}
accretion, accretion discs - stars: neutron - X-rays: binaries -
pulsars: individual: 2S 0114+650
\end{keywords}

\section{Introduction}

The high mass X-ray binary 2S 0114+650 has been shown to be a
rather unusual system. It exhibits properties consistent with both
Be and super-giant X-ray binaries \citep*{cra85}, with significant
broadband temporal and spectral variability over a wide range of
timescales \citep[e.g.][]{kon83,cra85}. The derived low X-ray
luminosity \citep[L$_x$ $\sim$1.1 $\times$ 10$^{36}$ erg s$^{-1}$
in the 3 -- 20 keV band for a distance of 7.2 kpc;][]{hal00}
implies that spherical accretion takes place via the stellar wind
of the donor star \citep{li99}. However, the HeII 4686 {\AA} line
(a common signature of the presence of an accretion disc) has been
observed weakly in emission on a few occasions (\citealt*{aab83};
\citealt{cra85,van89}), possibly linked to the presence of a small
transient accretion disc.

Coherent X-ray (and possibly faint optical) pulsations at a period
of $\sim$2.7 h have been shown to be a persistent feature of the
system \citep*{fin92,tay95} and have been variously attributed to
$\beta$ Cephei-type pulsations/tidal oscillations of the donor
star and the neutron star spin period. \citet*{soo06} analysed the
evolution of the pulse period in $\sim$9.5 yr of \emph{Rossi X-ray
Timing Explorer} (\emph{RXTE}) All Sky Monitor (ASM) data,
identifying two episodes of torque reversal. Their conclusions
support the scenario that 2S 0114+650 contains a super-slow
rotator -- one of the slowest spinning X-ray pulsars yet
discovered.

Variability at $\sim$11.6 d has been observed in both optical
\citep{cra85} and X-ray wavelengths \citep*{cor99}, consistent
with the orbital period of the system. \citet{cra85} were the
first to confirm the binary nature of 2S 0114+650 using radial
velocity measurements (with a period of 11.588 $\pm$ 0.003 d for
an eccentricity of 0.16 $\pm$ 0.07, and 11.591 $\pm$ 0.003 d for
an eccentricity of 0.0). Although they could not distinguish
between a circular and an eccentric orbit, they concluded that the
orbital motion was most likely eccentric as the system had
presumably gone through a supernova stage which should lead to a
perturbed orbit. \citet{cor99} analysed $\sim$2.5 yr of
\emph{RXTE} ASM observations of 2S 0114+650 between 1996 -- 1999,
identifying X-ray modulations close to both the optically derived
orbital period and the proposed pulse period. Fitting a sine wave
to the ASM light curve yielded a period of 11.63 $\pm$ 0.007 d.
The inconsistency between the optical and X-ray orbital period
values was explained as either due to variable contamination of
absorption lines by low equivalent width emission lines (resulting
from the X-ray irradiation of the donor star) or possibly
variability in the X-ray orbital modulation.

The determination of the orbital period was revisited recently in
both optical and X-ray wavelengths by \citet{gru07} and
\citet{wen06} respectively. \citet{gru07} performed a long-term
spectroscopic monitoring program of the H$\alpha$ emission in 2S
0114+650 and presented revised orbital elements with a period of
11.5983 $\pm$ 0.0006 d and a non-zero eccentricity of 0.18 $\pm$
0.05. \citet{wen06} re-analysed the ASM data and reported a new
value for the orbital period of 11.599 $\pm$ 0.005 d, consistent
with the new optical period.

A super-orbital modulation at 30.7 $\pm$ 0.1 d was reported by
\citet*{far06} from the analysis of 8.5 yr of ASM data.
\citet{wen06} subsequently reported a new improved value of 30.75
$\pm$ 0.03 d from re-analysis of the same data. \citet{soo06}
analysed the evolution of the super-orbital period, finding it to
be stable and concluding that 2S 0114+650 is only the fourth
reported system to show stable super-orbital variability.

The standard model invoked to explain super-orbital variability in
X-ray binaries is periodic obscuration or reflection via the
precession of a warped accretion disc \citep[e.g.][]{ogl01}. The
apparent lack of a persistent disc in this system thus raises
doubts about the validity of this model as an explanation for the
observed 30.7 d modulation. Recently, \citet*{kon06} presented a
model by which both the pulse and super-orbital modulations in
this system could be explained by tidally induced oscillations in
the donor star. However, the precise mechanisms by which these
oscillations may translate into a structured stellar wind are not
clear. In addition, super-orbital variability is predicted for
circular orbits while the eccentric orbit models predict strong
variations on orbital timescales associated with periastron
passage \citep{kon06}. These results are thus inconsistent with
the non-zero eccentricity recently derived by \citet{gru07}.

In this paper we present the results of a series of X-ray
observations of 2S 0114+650 performed with the Proportional
Counter Array \citep*[PCA; see][]{jah96} and High Energy X-ray
Timing Experiment \citep*[HEXTE; see][]{gru96} instruments onboard
the \emph{RXTE} satellite. The preliminary results of these
studies were published in \citet{far07}. The temporal and spectral
variability of this source was studied in detail over the pulse,
orbital and super-orbital timescales. In $\S$ 2 the data reduction
steps that were taken are explained. A description of the analysis
methods employed and the results obtained are provided in $\S$ 3.
A discussion of our results is provided in $\S$ 4, and $\S$ 5
presents the conclusions that we have drawn.

\section{Data Reduction}

Between 2005 May 15 -- 2005 June 14 (MJD 53505 -- 53535) and 2005
December 13 -- 2006 January 12 (MJD 53717 -- 53747) 2S 0114+650
was observed for $\sim$220 ks with the PCA and HEXTE instruments.
Two separate runs of 11 observations were made, with each
individual observation $\sim$3 hr in duration so as to cover one
complete cycle of the pulse period per observation. Each run
covered one super-orbital cycle. The two runs were spaced
approximately six months apart, with observations scheduled every
few days. Table~\ref{tabxobs} lists the 22 observations with the
date and exposure for each observation.

The Standard2f mode PCA data (16 s integration time, 128 energy
channels) were used for our timing and spectral analyses, as
time-resolution less than 16 s was not required. The HEXTE Archive
mode data (16 s integration time, 64 energy channels) were used
for our spectral analyses to investigate the high energy end of
the spectrum. The \textsc{ftools} v6.1 software package was used
for the reduction and analysis steps outlined below.

Filtering criteria were employed to exclude times when the
pointing direction was less than 10$^\circ$ from the Earth
horizon, the offset from the target direction was greater than
0.02$^\circ$ or the electron contamination fraction exceeded 0.1.
Data obtained within 30 min of the spacecraft passage through the
South Atlantic Anomaly (SAA) were also excluded to remove the
effect of residual high-background levels due to induced
radioactivity.

Spectra were extracted from all anode layers for each Proportional
Counter Unit (PCU) in the PCA from all data sets, excluding PCU0
due to the increase in the background that occurred therein with
the loss of the propane
layer\footnote{http://universe.gsfc.nasa.gov/xrays/programs/rxte/pca/doc/\\bkg/bkg-2002/}.
Spectra were also extracted for both HEXTE clusters (cluster 0 and
1) from all data sets. Due to the intermittent failure of the
rocking mechanism in cluster 0 during our second set of
observations\footnote{http://heasarc.gsfc.nasa.gov/docs/xte/whatsnew/big.html},
background data for this cluster are not available for
observations 12 -- 19. As a result, cluster 0 data from these
observations are not included in our spectral analyses.

\begin{table}
\begin{center}
\caption{$\emph{RXTE}$ pointed observations of 2S
0114+650.}\label{tabxobs}
\begin{tabular}{cccccc}
\hline Obs. & Date &  Exposure & Obs.
 & Date &  Exposure\\
& (MJD) & (s) & & (MJD) & (s)\\
\hline
1 & 53506 & 7,104 & 12 & 53717 & 7,216 \\
2 & 53509 & 7,488 & 13 & 53721 & 10,176 \\
3 & 53513 & 5,856 & 14 & 53723 & 8,608 \\
4 & 53515 & 7,408 & 15 & 53726 & 8,064 \\
5 & 53517 & 9,120 & 16 & 53729 & 8,336 \\
6 & 53520 & 9,616 & 17 & 53732 & 1,632 \\
7 & 53523 & 8,960 & 18 & 53735 & 7,888 \\
8 & 53526 & 6,640 & 19 & 53738 & 7,616 \\
9 & 53529 & 7,152 & 20 & 53741 & 7,664 \\
10 & 53533 & 7,056 & 21 & 53745 & 6,752 \\
11 & 53535 & 7,408 & 22 & 53747 & 8,592 \\
\hline
\end{tabular}

\end{center}
\end{table}

Three light curves covering the 2 -- 9 keV (A), 9 -- 20 keV (B)
and 20 -- 40 keV (C) energy ranges were extracted from all layers
of PCU1, PCU2, PCU3 and PCU4 separately. The energy boundaries
were chosen in line with those of the standard product light
curves provided with the data. As the average count rate before
background subtraction was less than 40 counts s$^{-1}$, the 2005
November 28 mission-long faint source model was used to generate
PCA background spectra and light curves. HEXTE background spectra
for each cluster were extracted from the ``off-source'' data, when
the cluster was aimed in a direction offset from the pointing
direction. PCA response matrices were generated for each
individual source spectrum. The 2000 May 26 HEXTE response and
ancillary response matrices were used, as the response of each
cluster has not changed since that time. As the observed count
rates were relatively low, a deadtime correction was not applied
to either the spectra or light curves.

The individual PCU spectra were then combined (source and
background, separately) to maximise the signal-to-noise ratio. The
PCA response matrices were also combined, weighted by the exposure
values for each
PCU\footnote{http://heasarc.gsfc.nasa.gov/docs/xte/recipes/pcu$\_$combine.html}.
For the analysis of the complete spectrum covering all
observations, a 1$\%$ systematic
error\footnote{http://heasarc.gsfc.nasa.gov/docs/xte/ftools/xtefaq$\_$answers.html}
was added to the resulting spectrum to account for uncertainties
in the response matrix such as the feature present at $\sim$5 keV
due to the xenon L-edges. For the phase resolved spectral analyses
no systematic error was applied as our aim was to constrain the
spectral parameters to the highest level possible, particularly in
the low energy range of the spectrum. Instead, the features around
$\sim$5 keV were modelled using a multiplicative absorption edge
component during the spectral fitting.

At energies below 3 and above 30 keV there were poor counting
statistics and so these data were excluded from the PCA spectrum.
The observed scatter in the HEXTE spectra is of the level expected
for systematic uncertainties in the background \citep{rot98}, and
so a 1$\%$ systematic uncertainty was added to its PHA channel
data. Energies less than the 17 keV low energy threshold were also
excluded in the HEXTE spectra. The low flux of 2S 0114+650 at high
energies resulted in poor statistics above 50 keV in the HEXTE
spectra. As such, energies above this range were excluded for our
analyses. The individual PCU light curves were averaged together
to give source and background light curves with the best
signal-to-noise ratio. Barycentric correction was applied and
background subtraction performed.

\section{Data Analysis}

\begin{table}
\begin{center}
\caption[\textsc{xspec} spectral models applied to the combined
PCA and HEXTE spectra]{\textsc{xspec} spectral models applied to
the combined PCA and HEXTE spectra. A photo-electric absorption
component (WABS in \textsc{xspec}) was included for all
models.}\label{tab51}
\begin{tabular}{lcc}
\hline Model & \textsc{xspec} Models &$\chi$$^2$/DOF\\
\hline
Power law + HE cut-off & POW*HIGHE & 28.8/80 \\
Black body + bremsstrahlung & BBODY+BREMS & 29.6/80\\
Cut-off power law & cut-off & 32.5/81\\
Power law + bremsstrahlung & POW+BREMS & 33.2/80\\
Comptonisation & COMPTT & 56.9/79\\
Thermal bremsstrahlung & BREMS & 63.1/82\\
Comptonisation & COMPST & 94.7/81\\
Black body & BBODY & 3890/82\\
\hline
\end{tabular}
\end{center}
\end{table}

\begin{figure*}
\begin{center}
\includegraphics[width=13cm]{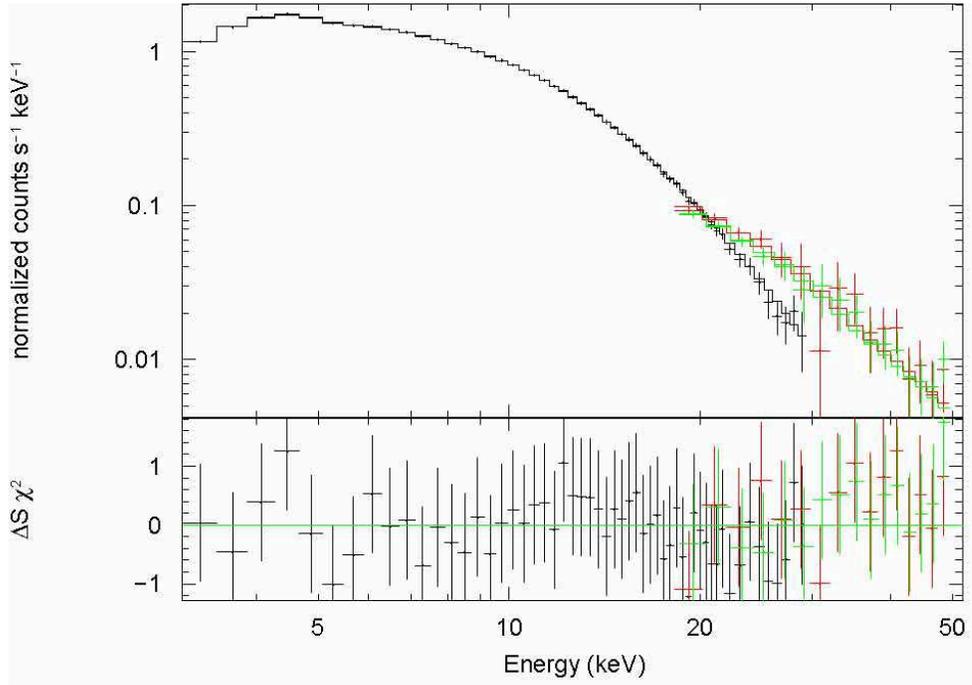}
\caption[3 -- 30 keV PCA and 17 -- 50 keV HEXTE cluster 0 $\&$ 1
spectra from all data.]{3 -- 30 keV PCA (black) and 17 -- 50 keV
HEXTE cluster 0 (red) $\&$ cluster 1 (green) spectra from all
data. The residuals to the fit are shown in the bottom panel, in
units of sigma ($\Delta$S) $\chi^2$.}\label{figAll}
\end{center}
\end{figure*}

\begin{table}
\begin{center}
\caption[Best fit X-ray spectral parameters of 2S 0114+650]{Best
fit X-ray spectral parameters of 2S 0114+650, achieved using an
absorbed power law with a high energy exponential cut-off. The
flux quoted is not corrected for absorption.}\label{tab52}
\begin{tabular}{lcc}
\hline Spectral Parameter & Value & Units\\
\hline Column Density & (3.2 $^{+ 0.9}_{- 0.8}$) $\times$ 10$^{22}$ & atoms cm$^{-2}$\\
Photon Index & 1.1 $\pm$ 0.1&\\
Cut-off Energy & 6.0 $\pm$ 0.7 & keV\\
Folding Energy & 15 $^{+ 3}_{- 2}$ & keV\\
Flux (3 -- 50 keV) & 2.3 $\times$ 10$^{-10}$ & ergs cm$^{-2}$ s$^{-1}$\\
\hline
\end{tabular}
\end{center}
\end{table}

\begin{figure*}
\begin{center}
\includegraphics[width=14.5cm]{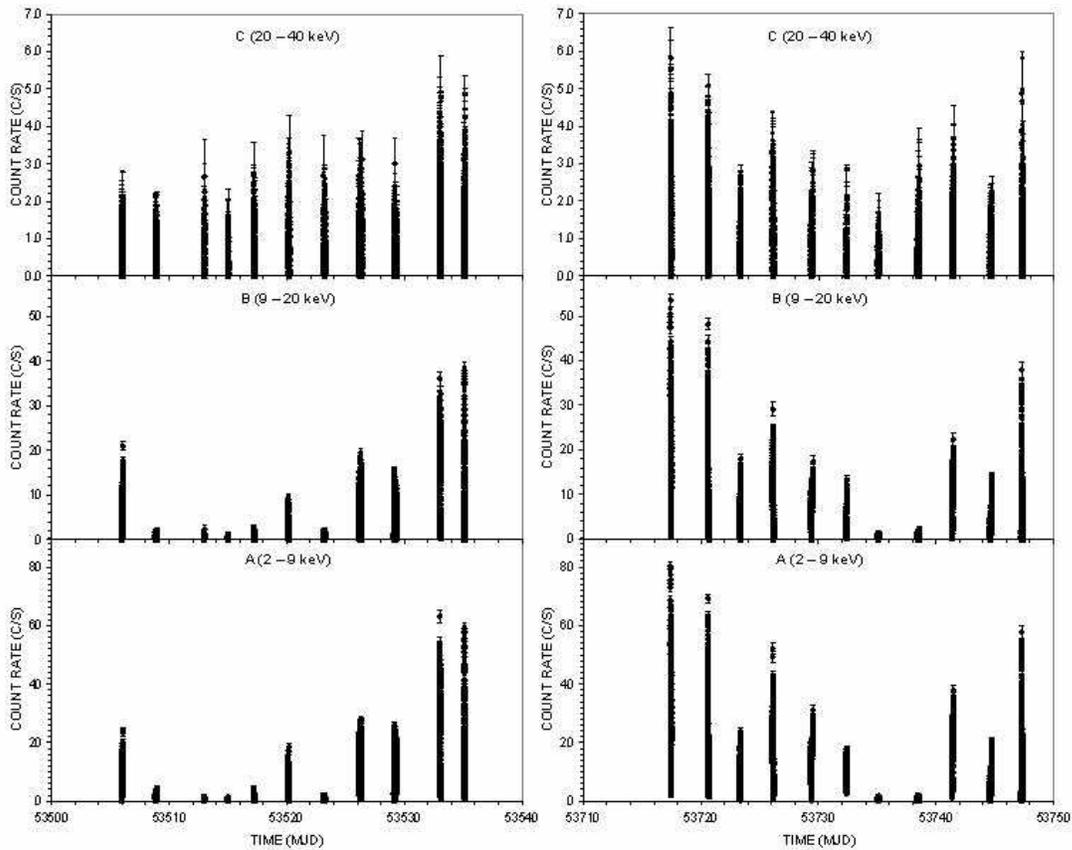}
\caption{PCA light curves of observations 1 -- 11 (left) and 12 --
22 (right): 2 -- 9 keV (bottom panels), 9 -- 20 keV (middle
panels) and 20 -- 40 keV (top panels).}\label{lc1}
\end{center}
\end{figure*}

\begin{figure}
\begin{center}
\includegraphics[width=8cm]{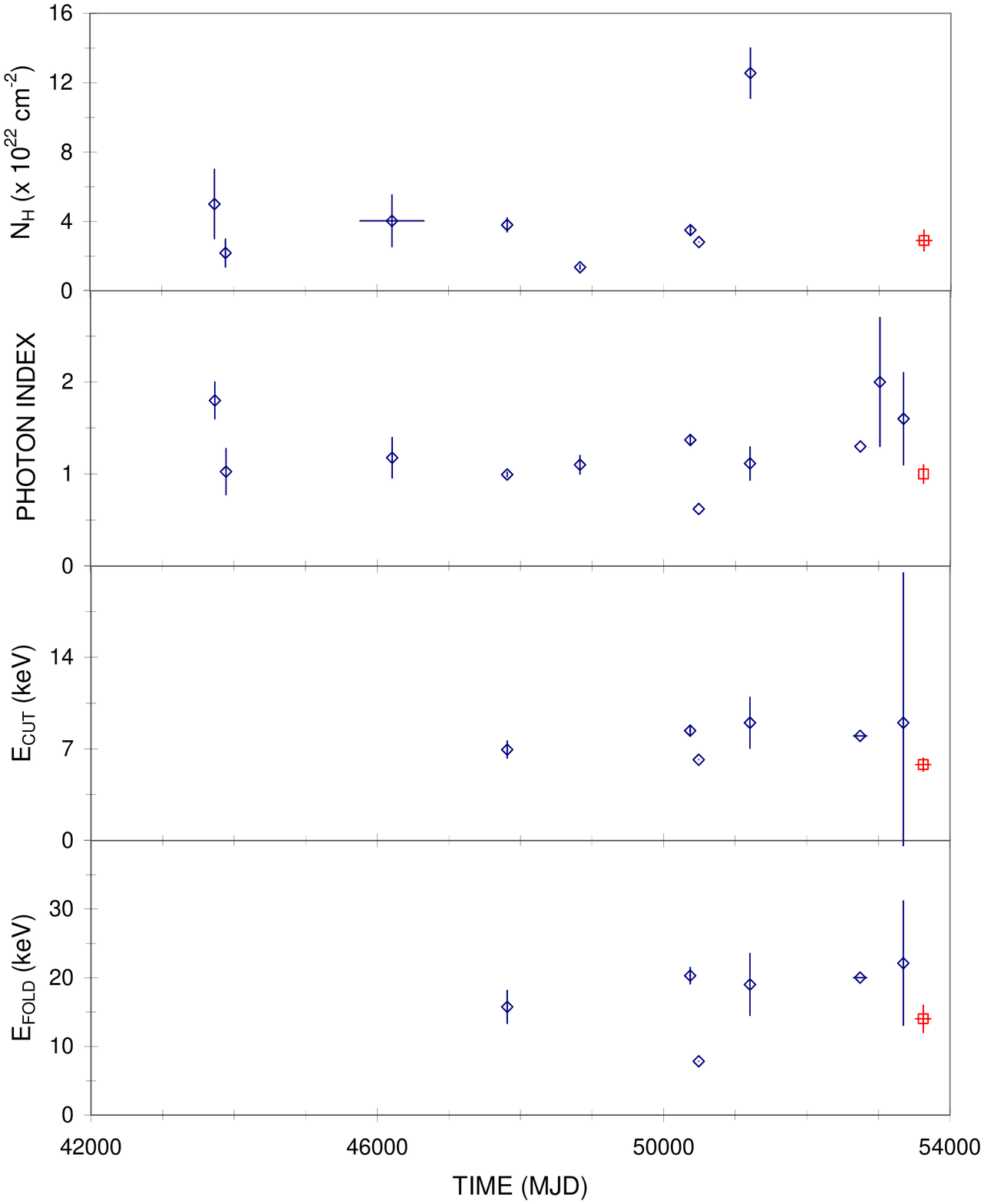}
\caption[Best fit spectral parameters compared against values from
previous observations]{Best fit spectral parameters (rightmost red
squares) compared against values from previous observations
\citep[blue
diamonds;][]{yam90,apa91,fin94,hal00,kon03,bon05,fil05,den06,mas06,muk06}.
From top to bottom: neutral hydrogen column density, power law
photon index, cut-off energy and folding energy.}\label{specomp}
\end{center}
\end{figure}

After the data reduction steps outlined above, spectral models
were fitted to the combined PCA and HEXTE (cluster 0 and 1,
separately) spectra in the 3 -- 50 keV energy range using
\textsc{xspec} v12.3. Eight models commonly applied to X-ray
binaries were tested, with the resulting $\chi^2$ values and
degrees of freedom presented in Table~\ref{tab51}. The best fit
model was found to be an absorbed power law with a high energy
(HE) exponential cut-off, with parameter values consistent with
the previously obtained values listed in the literature
\citep*{yam90,apa91,fin94,hal00,kon03,bon05,fil05,den06,mas06,muk06}.

The reduced $\chi^2$ values obtained by fitting 5 of the 7
remaining models were $<$ 1, also indicating acceptable fits. Many
of the models (such as the CUTOFF, COMPTT and COMPST) have the
same general shape as the empirical power law + HE cut-off model,
but attempt to provide a better physical description of the
underlying emission mechanisms. The other models (the black body
and thermal bremsstrahlung) provided poor fits on their own, and
required more complicated combinations in order to adequately
model the overall spectrum. For these more complicated models we
applied the F-test in \textsc{xspec} in order to justify the
addition of the extra components. The addition of a soft black
body component to the thermal bremsstrahlung model produced an
F-statistic of 12.9 and a probability of 5 $\times$ 10$^{-11}$,
indicating a significant improvement in the fit. The addition of a
power law component to the bremsstrahlung model \citep[a
combination suggested by][]{den06} yielded an F-statistic of 10.3
and a probability of 4 $\times$ 10$^{-9}$, also indicating an
improved fit.

While we cannot discriminate between 6 of the models from spectral
fitting alone, we have adopted the absorbed power law with a HE
cut-off model (WABS*POWER*HIGHE) as it is commonly used to
empirically describe the shape of the spectra of accreting X-ray
pulsars, and has been used in previous spectral analyses of 2S
0114+650 \citep[e.g.][]{hal00}.

Inclusion of an Fe K$\alpha$ Gaussian emission line at 6.4 keV did
not improve the fit significantly in the overall spectrum.
Similarly, no evidence was found of the $\sim$22 keV cyclotron
resonance scattering feature (CRSF) tentatively identified in
\emph{INTEGRAL} data by \citet{bon05}. Our results thus support
the findings of \citet{mas06} and \citet{den06}, who failed to
detect any evidence of cyclotron features in the spectrum of 2S
0114+650 using data from the \emph{BeppoSAX} and \emph{INTEGRAL}
satellites respectively. Upper limits at the 95$\%$ confidence
level for the equivalent width of the iron line and the depth of
the 22 keV cyclotron feature of 55 eV and 0.3 respectively were
obtained.

Figure~\ref{figAll} shows the combined PCA and HEXTE spectra from
all observations fitted with the best fit model described above.
Figure~\ref{lc1} shows the PCA A, B and C light curves from both
observation runs covering MJD 53505 -- 53535 and MJD 53717 --
53747. Figure~\ref{specomp} compares the best fit spectral
parameters with values from previous observations listed in the
literature. The PCA A band light curves for all 22 individual
observations are included in Appendix A.

The $\sim$2.7 h pulse period can clearly be seen in most of the
light curves, appearing as a smooth near-sinusoidal modulation
(see Figures \ref{1lc} -- \ref{4lc}). Intermittent flaring
episodes can be seen during many of the observations occurring in
phase with the maximum of the pulse period (see e.g. the light
curve from observation 13 in Figure \ref{3lc} in the Appendix).
These flaring episodes appear to be pseudo-periodic, possibly
indicative of the 850 s, 894 s, or $\sim$2,000 s periods reported
by \citet{yam90}, \citet{kon83} and \citet*{apa91} respectively.
In addition to these reports of periodic modulation, \citet{mas06}
reported observing ``X-ray flickering'' on timescales of minutes
to hours, which they speculated were a result of random
inhomogeneities in the accretion flow onto the neutron star.

Two distinct flares were seen to occur during observation 6
(Figure~\ref{1lc}) despite the otherwise low flux from the source
at this time. These flares are predominantly present at low
energies, indicating a source within the PCA field of view as the
origin and ruling out gamma ray bursts as the cause. \citet{mas06}
suggested that similar flares observed in the \emph{BeppoSAX} data
were due to anisotropies in the density of the stellar wind (and
thus the accreting material).

The light curve of observation 22 (Figure~\ref{4lc}) exhibits an
episode of very low flux during the first hour of the observation,
after which the pulse period materialises with large amplitude. As
mentioned earlier, all data where the pointing offset was greater
than 0.02$^{\circ}$ from the target source were discarded. As such
this low-flux episode is not due to the spacecraft slewing into
position.

Spectral and timing analyses were performed over the pulse,
orbital and super-orbital period timescales. Analyses of the
flaring episodes observed at the peak of the pulse period were
also performed. These analyses are described below.

\subsection{Pulse Variability}

\begin{figure}
\begin{center}
\includegraphics[width=8cm]{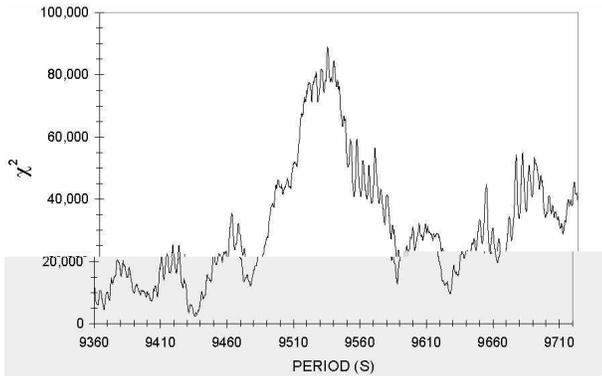}
\caption[Epoch folding $\chi$$^2$ search for the pulse period in
the 2 -- 9 keV PCA light curve covering the first observation
run]{Epoch folding $\chi$$^2$ search for the pulse period in the 2
-- 9 keV PCA light curve covering the first observation run (MJD
53505 -- 53535).}\label{p1aefs}
\end{center}
\end{figure}

\begin{figure}
\begin{center}
\includegraphics[width=8cm]{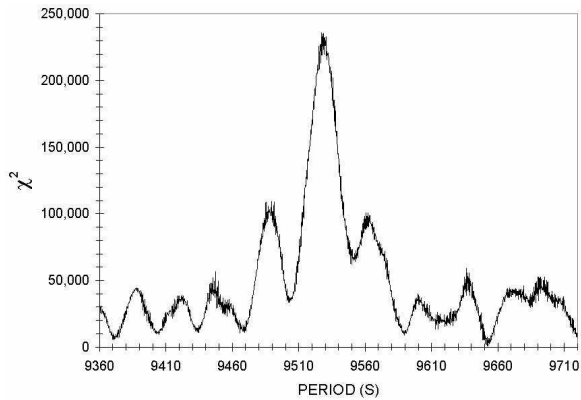}
\caption[Epoch folding $\chi$$^2$ search for the pulse period in
the 2 -- 9 keV PCA light curve covering the second observation
run]{Epoch folding $\chi$$^2$ search for the pulse period in the 2
-- 9 keV PCA light curve covering the second observation run (MJD
53717 -- 53747).}\label{p2aefs}
\end{center}
\end{figure}

\begin{figure}
\begin{center}
\includegraphics[width=8cm]{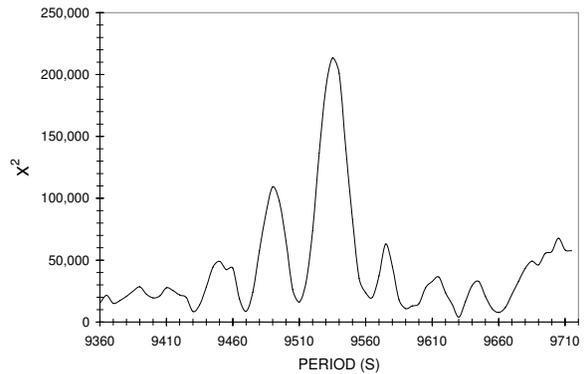}
\caption[Epoch folding $\chi$$^2$ search for the pulse period in
the 2 -- 9 keV PCA light curve covering MJD 53717 -- 53747]{Epoch
folding $\chi$$^2$ search for the pulse period in the 2 -- 9 keV
PCA light curve covering all observations (MJD 53505 --
53747).}\label{ptotal}
\end{center}
\end{figure}

An epoch folding $\chi$$^2$ search was performed on the PCA 2 -- 9
keV light curve using the \textsc{ftools} task \textsc{efsearch}
in the 9,360 -- 9,720 s range with 0.36 s resolution. Light curves
from the two observation runs (observations 1 -- 11 and 12 -- 22)
were analysed separately to determine whether the pulse period
varied significantly over the 6 month period separating the
observation runs. The values for the pulse period were found to be
2.65 $\pm$ 0.01 h and 2.644 $\pm$ 0.003 h respectively, indicating
the pulse period did not vary significantly over the 6 month
period between the observation runs. The \textsc{efsearch}
analysis of the light curve covering both runs of observations
(with a reduced resolution so as to ``smooth out'' aliasing
effects) yielded a best fit period of 2.650 $\pm$ 0.003 h FWHM.
Figures~\ref{p1aefs}, \ref{p2aefs} and \ref{ptotal} show the
$\chi$$^2$ vs period plots from the \textsc{efsearch} analysis.

\begin{figure}
\begin{center}
\includegraphics[width=8cm]{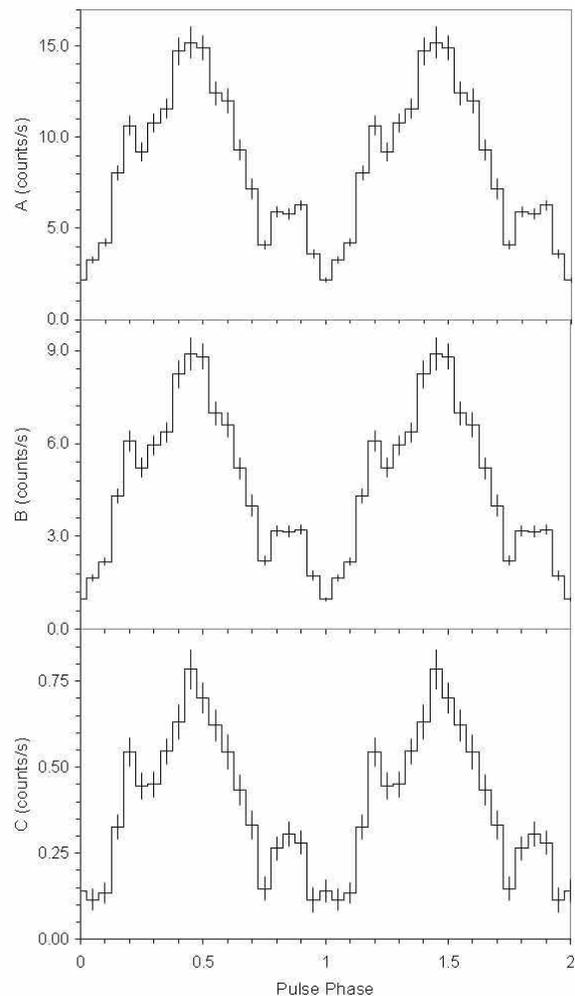}
\caption[PCA Standard2 A (top), B (middle) and C (bottom) light
curves of all observations folded over the 2.65 h pulse
period.]{PCA Standard2 A (top), B (middle) and C (bottom) light
curves of all observations folded over the 2.65 h pulse period,
with error bars determined from the standard error. Two cycles are
shown for clarity.}\label{pulseA}
\end{center}
\end{figure}

In order to investigate the profile of the pulse period the A, B
and C (Figure~\ref{lc1}) PCA light curves covering all
observations were folded over the best fit 2.65 h period using the
task \textsc{efold}, with phase 0 set at the flux minimum at MJD
53504.912 and with the errors being the standard error for each
phase bin. The pulse profile was found to be approximately
sinusoidal with a short ``spike'' around phase 0.2, a larger
feature appearing around phase 0.85, and little variation between
energy bands (Figure \ref{pulseA}).

If 2S 0114+650 contains a super-slow X-ray pulsar with a 2.65 h
spin period, the pulse period should exhibit Doppler shifting over
the 11.6 d orbital period by $\sim$16 s. Likewise, if the 30.7 d
super-orbital period is related to the orbital period of a second
compact object in the system \citep[as postulated by][]{far06},
the pulse period should exhibit Doppler shifting over this
timescale by $\sim$12 s (the orbital angle of inclination
permitting). As each of the 22 observations cover less than two
pulse cycles, we were unable to measure the pulse arrival times in
each data set to the required level of accuracy to constrain any
variability over the orbital or super-orbital timescales.
Additionally, it is unclear whether it is possible to measure the
pulse period to the required level of accuracy ($\sim$0.1$\%$ of
the pulse period) with further observations.

To investigate the spectral variability over the pulse period, we
extracted ten phase resolved spectra separated by 0.1 phase bins
over the 2.65 h pulse period using the same method outlined
earlier. For these analyses, only the PCA 3 -- 30 keV data were
utilised due to the poor statistics in the HEXTE data resulting
from the relatively short exposures in each phase bin. As the
objective of these analyses was to constrain the spectral
parameters in each phase bin as tightly as possible, the 1$\%$
systematic error that was applied to the combined spectrum from
all data was removed. The incorrect modelling of the xenon L-edge
around $\sim$5 keV was instead accounted for by an additional
absorption component for these analyses, as outlined in \S 2.

\begin{figure}
\begin{center}
\includegraphics[width=\columnwidth]{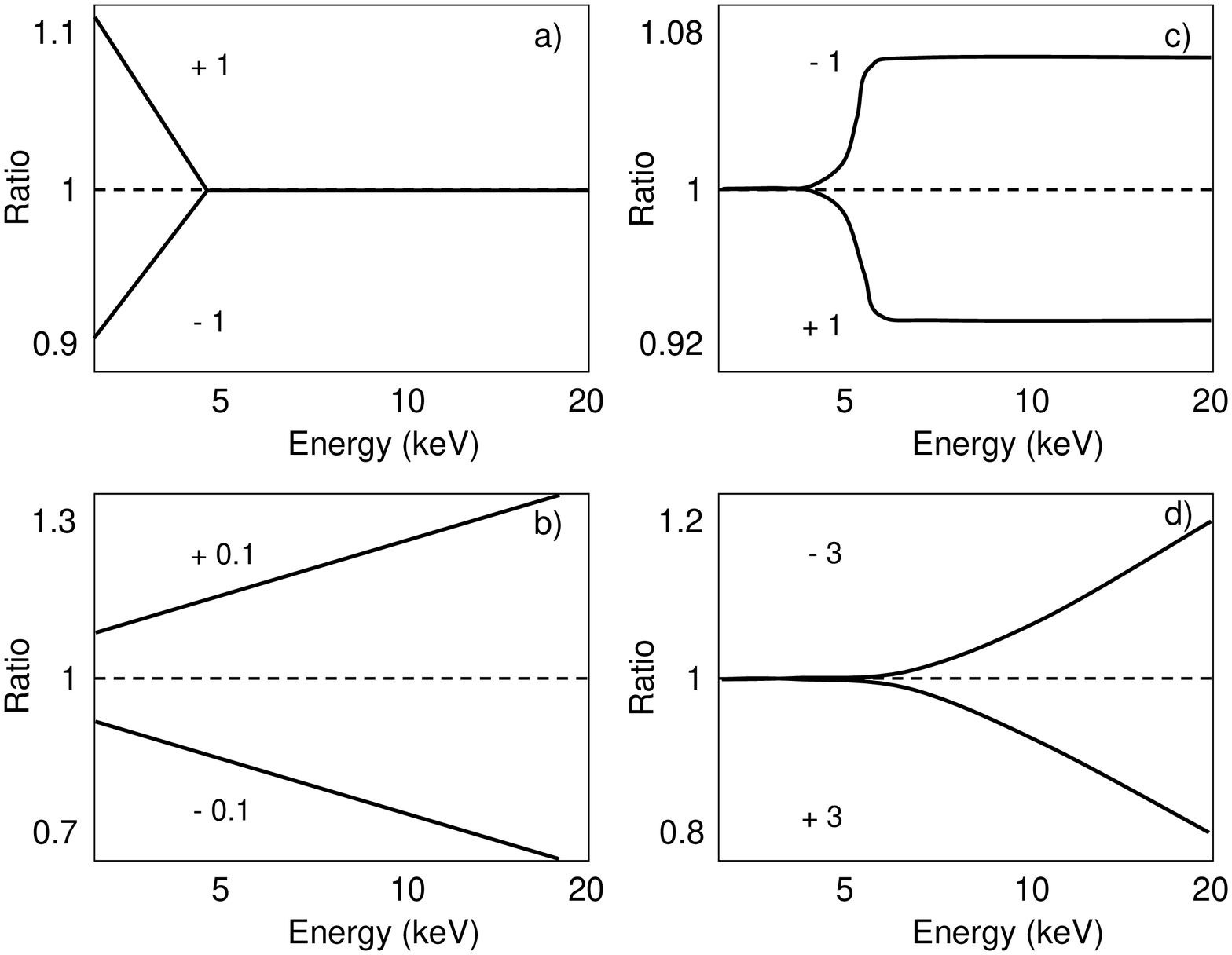}
\caption{The effects of varying spectral parameters on the ratio
between the best fit model and the phase resolved spectra from
simulations: a) varying the column density by $\pm$ 1 $\times$
10$^{22}$ cm$^{-2}$, b) varying the photon index by $\pm$ 0.1, c)
varying the cut-off energy by $\pm$ 1 keV, and d) varying the
folding energy by $\pm$ 3 keV.}\label{specvar}
\end{center}
\end{figure}

Comparing the individual phase resolved spectra with the best fit
model obtained using the overall spectrum in Figure \ref{figAll}
allows for the identification of any energy specific spectral
variability. In order to determine the source of any spectral
variability, we first performed simulations using the overall
spectrum. The spectral parameters of the best fit model were
modified and the ratio between the modified model and the overall
spectrum were calculated. As can be seen in Figure \ref{specvar},
varying each of the spectral parameters in turn resulted in quite
specific patterns within the ratio plots. Varying the power law
normalisation parameter (not shown in Figure \ref{specvar}) simply
shifted the ratio values up or down, while keeping the slope of
the ratios flat. Varying the column density modified the ratios
only below $\sim$5 keV, while varying the parameters of the high
energy cut-off (i.e. the cut-off energy and folding energy
parameters) modified the ratios above $\sim$5 keV. Varying the
photon index modified the slope of the ratio lines, while varying
all parameters simultaneously produced a combination of all above
mentioned effects.

\begin{figure*}
\begin{center}
\includegraphics[width=18cm]{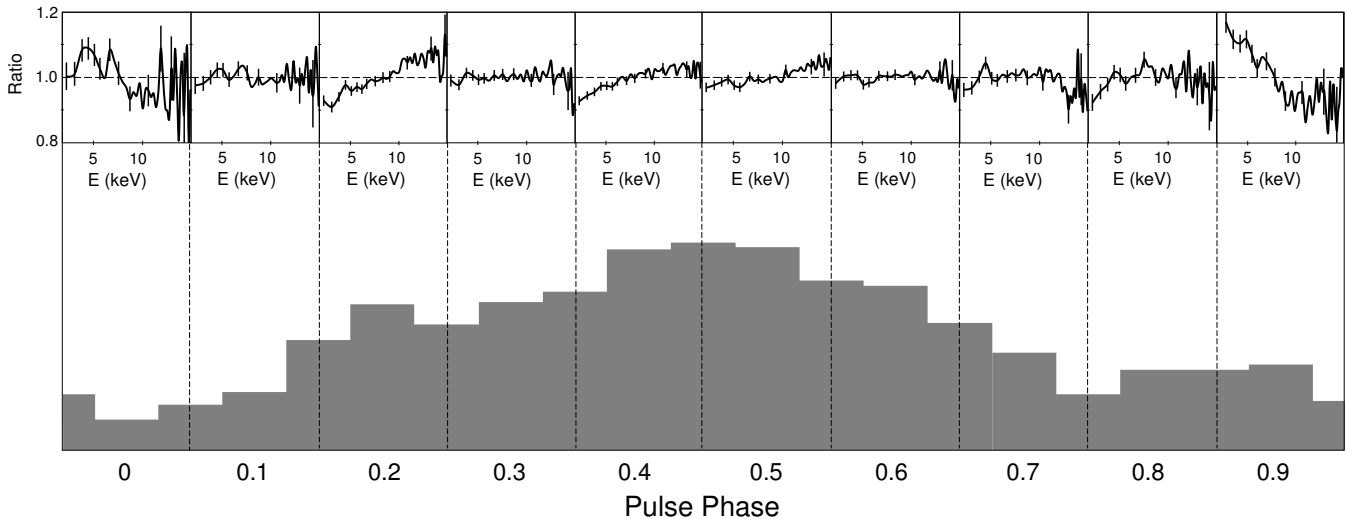}
\caption{Ratios between the pulse phase resolved spectra and the
best fit model from the overall spectrum (top inset figures)
plotted against the folded pulse profile (histogram). For clarity
only representative error bars are plotted on the ratio
plots.}\label{prat}
\end{center}
\end{figure*}

The ratios between each of the pulse phase resolved spectra and
the best-fit model obtained from the overall spectrum were then
plotted in order to identify any spectral variability over the
pulse period, and to identify the energy ranges in which the
spectra were variable. The results are plotted in Figure
\ref{prat}. For clarity the spectral model was normalised for each
of the spectra through the addition of a constant parameter,
removing the offset produced by the differing power law
normalisation.

The ratio plots shown in Figure \ref{prat} indicate significant
spectral variability over the pulse period. Comparison with the
results of the simulations presented in Figure \ref{specvar}
indicates likely variability in the slope of the power law and in
the column density parameters, with no evidence for variability in
the high energy cut-off parameters. Additional features in the
ratios around $\sim$6 keV indicate possible flux dependant
variability in the Fe K$\alpha$ emission.

Based on this comparison, we fit each of the pulse phase resolved
spectra simultaneously with an absorbed power law model with a
high energy cut-off, with the cut-off energy and folding energy
parameters tied together, but with the column density and power
law parameters free to vary independently. In this method the
cut-off parameters are not fixed, but are linked so that while
they were free to vary overall, the values were the same for each
of the phase resolved spectra.

Significant features in the residuals were observed around $\sim$5
keV and $\sim$6 keV, which we attribute to the incomplete
modelling of the xenon L-edges and to fluorescent Fe K$\alpha$
emission. Additional model components comprising an absorption
edge around $\sim$5 keV and a Gaussian emission line at 6.4 keV
(with the energy frozen at 6.4 keV but the sigma and normalization
parameters free to vary between the spectra) were thus included.
While we cannot determine whether the Fe line is a result of
fluorescent emission from 2S 0114+650 or from the diffuse galactic
background, similar features have been seen in the spectrum before
\citep[e.g.][]{yam90,apa91,ebi97,hal00,mas06,muk06} and so we
assume that the observed line is related to the source. The
reduced $\chi^2$ statistic obtained from this fitting process was
0.531 with 478 degrees of freedom, indicating a very good fit.

\begin{figure}
\begin{center}
\includegraphics[width=\columnwidth]{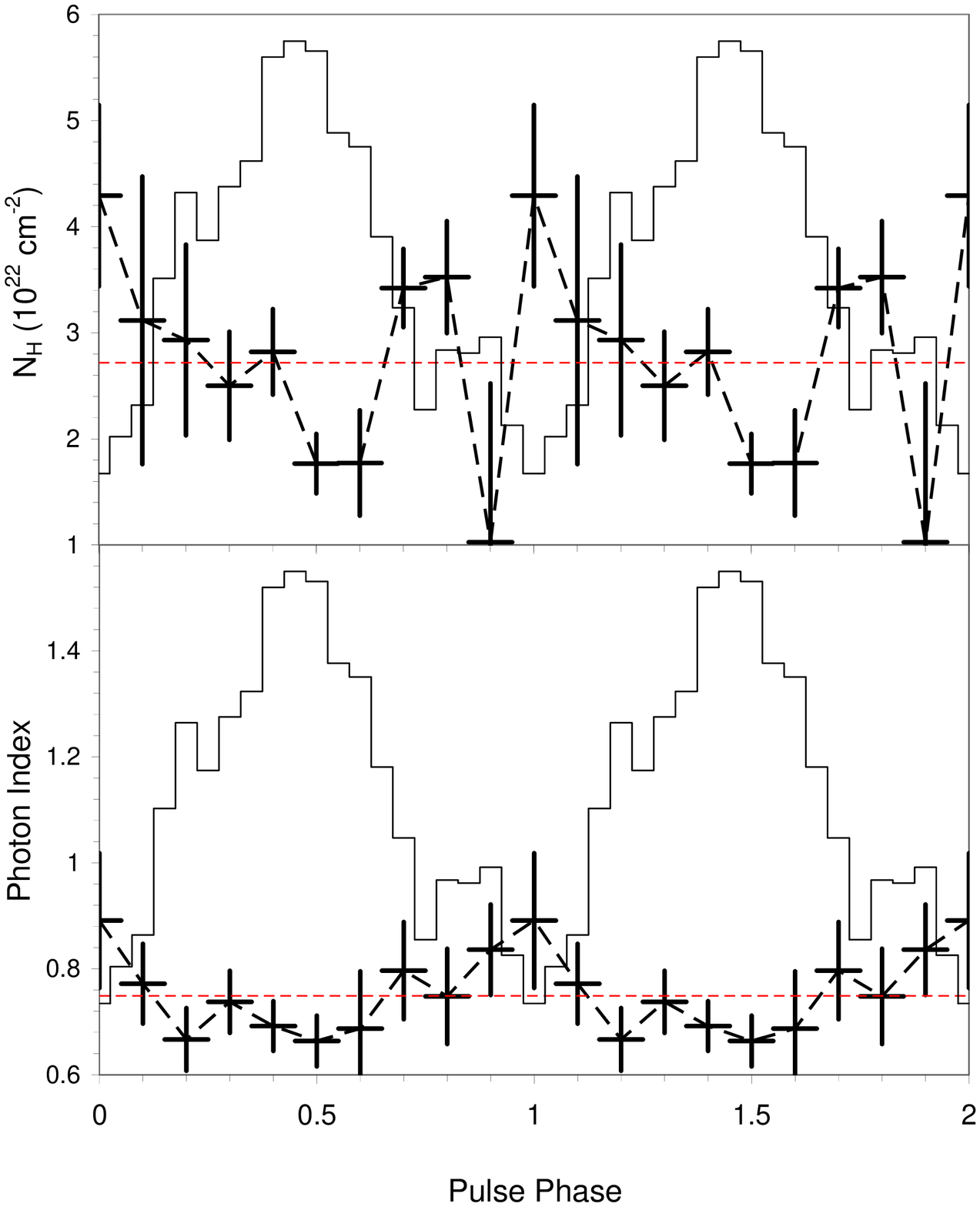}
\caption{Variability of the column density (top) and power law
photon index (bottom) over the pulse period, with error bars
representing 90$\%$ confidence intervals. The histogram represents
the \textit{RXTE} PCA 2 -- 9 keV light curve folded over the best
fit 2.65 h pulse period. The red dashed lines indicate the mean
column density and photon index.}\label{specpuls1}
\end{center}
\end{figure}

The column density and power law photon index parameters were then
determined and plotted against pulse phase and compared with the
PCA 2 -- 9 keV light curve folded over the same period (Figure
\ref{specpuls1}). The reduced $\chi^2$ value for the hypothesis of
parameter constancy was then calculated, with values of 1.41 (null
hypothesis probability of 18$\%$) and 1.23 (null hypothesis
probability of 27$\%$) obtained for the column density and photon
index respectively, providing marginal evidence of variability in
both parameters over the pulse period. The linear correlation
coefficient was also calculated for each parameter compared to the
folded pulse profiles. The value obtained for the column density
was -0.22, indicating a weak anti-correlation, while the value
obtained for the photon index was -0.82, indicating a strong
anti-correlation. While the addition of an Fe K$\alpha$ emission
line was required for the best fit, we were unable to constrain
the equivalent width of the line sufficiently in order to
determine whether it is variable over the pulse period. The pulse
phase resolved spectra are shown in Figures \ref{pulsespec1} and
\ref{pulsespec2} in the Appendix.

\subsection{Orbital Variability}

\begin{figure}
\begin{center}
\includegraphics[width=8cm]{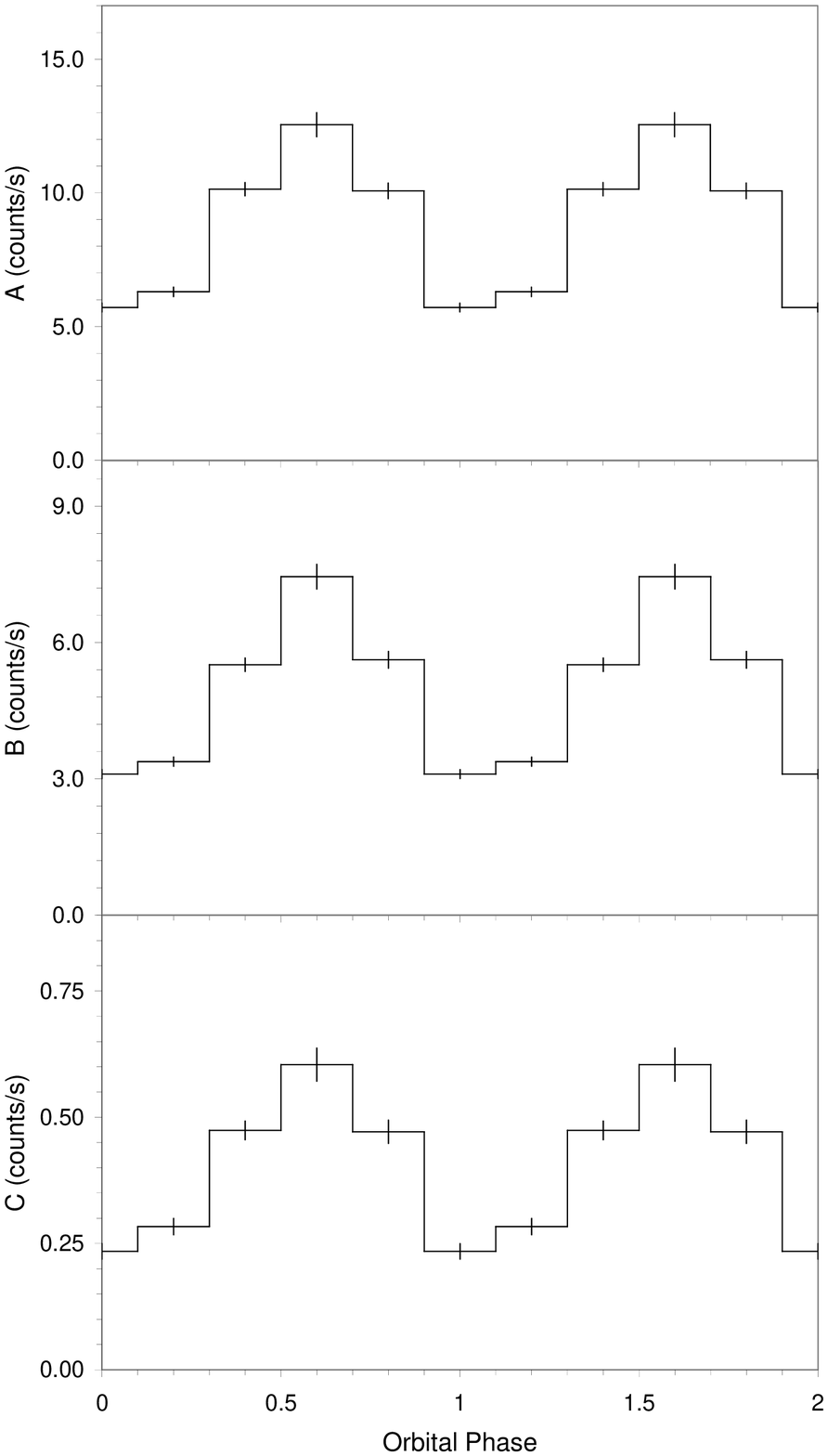}
\caption[PCA Standard2 A (top), B (middle) and C (bottom) light
curves of all observations folded over the 11.99 d orbital
period.]{PCA Standard2 A (top), B (middle) and C (bottom) light
curves of all observations folded over the 11.599 d orbital
period, with error bars determined from the standard
error.}\label{orbA}
\end{center}
\end{figure}

Approximately 6 orbital cycles were sampled during the course of
our $\emph{RXTE}$ observations. Due to the limited number of
cycles and relatively poor phase coverage, the value of the
orbital period could not be accurately constrained. As such, the
value of 11.599 d obtained by \citet{wen06} was used for our
analyses. The A, B and C (Figure~\ref{orbA}) PCA light curves
covering all observations were folded over 11.599 d, with phase 0
set at the flux minimum at MJD 53503.7. The profile of the
modulation does not appear to vary significantly with energy.

\begin{figure}
\begin{center}
\includegraphics[width=\columnwidth]{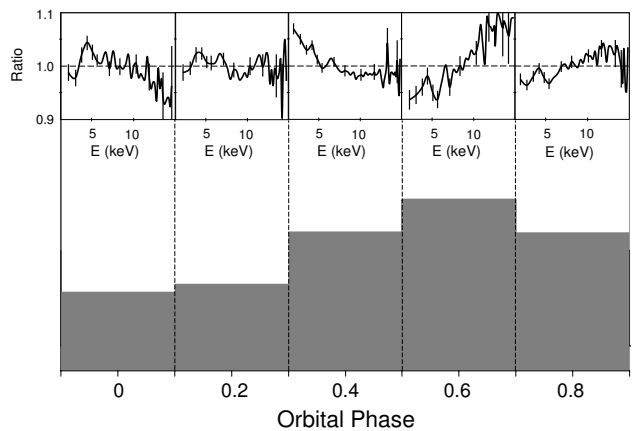}
\caption{Ratios between the orbital phase resolved spectra and the
best fit model from the overall spectrum (top inset figures)
plotted against the folded orbital profile
(histogram).}\label{orat}
\end{center}
\end{figure}

Using the same method outlined above for the pulse period, we
extracted five phase resolved spectra separated by 0.2 phase bins
from the PCA data over the 11.599 d orbital period. The ratios
between each of the orbital phase resolved spectra and the
best-fit model obtained from the overall spectrum were then
plotted in order to identify any spectral variability. The results
are plotted in Figure \ref{orat}.

The ratio plots shown in Figure \ref{orat} indicate significant
spectral variability over the orbital period similar to that seen
over the pulse period. Comparison with the results of the
simulations presented in Figure \ref{specvar} indicates likely
variability in the slope of the power law and in the column
density parameters, with no evidence for variability in the high
energy cut-off parameters. Features in the ratios around $\sim$6
keV are present similar to that seen in the pulse ratio plots,
again indicating possible variability in the Fe K$\alpha$
emission.

Based on this comparison, the same absorbed power law with high
energy cut-off model was fitted to each of the orbital phase
resolved spectra and the spectral parameters measured. As with the
pulse phase resolved spectral analyses, the cut-off energy and
folding energy parameters were tied together, but with the column
density and power law parameters free to vary independently.
Additionally, an absorption edge around $\sim$5 keV and a Gaussian
emission line at 6.4 keV were included to account for residuals
observed in the low energy range of the spectra. The reduced
$\chi^2$ statistic obtained from this fitting process was 0.675
with 239 degrees of freedom.

\begin{figure}
\begin{center}
\includegraphics[width=\columnwidth]{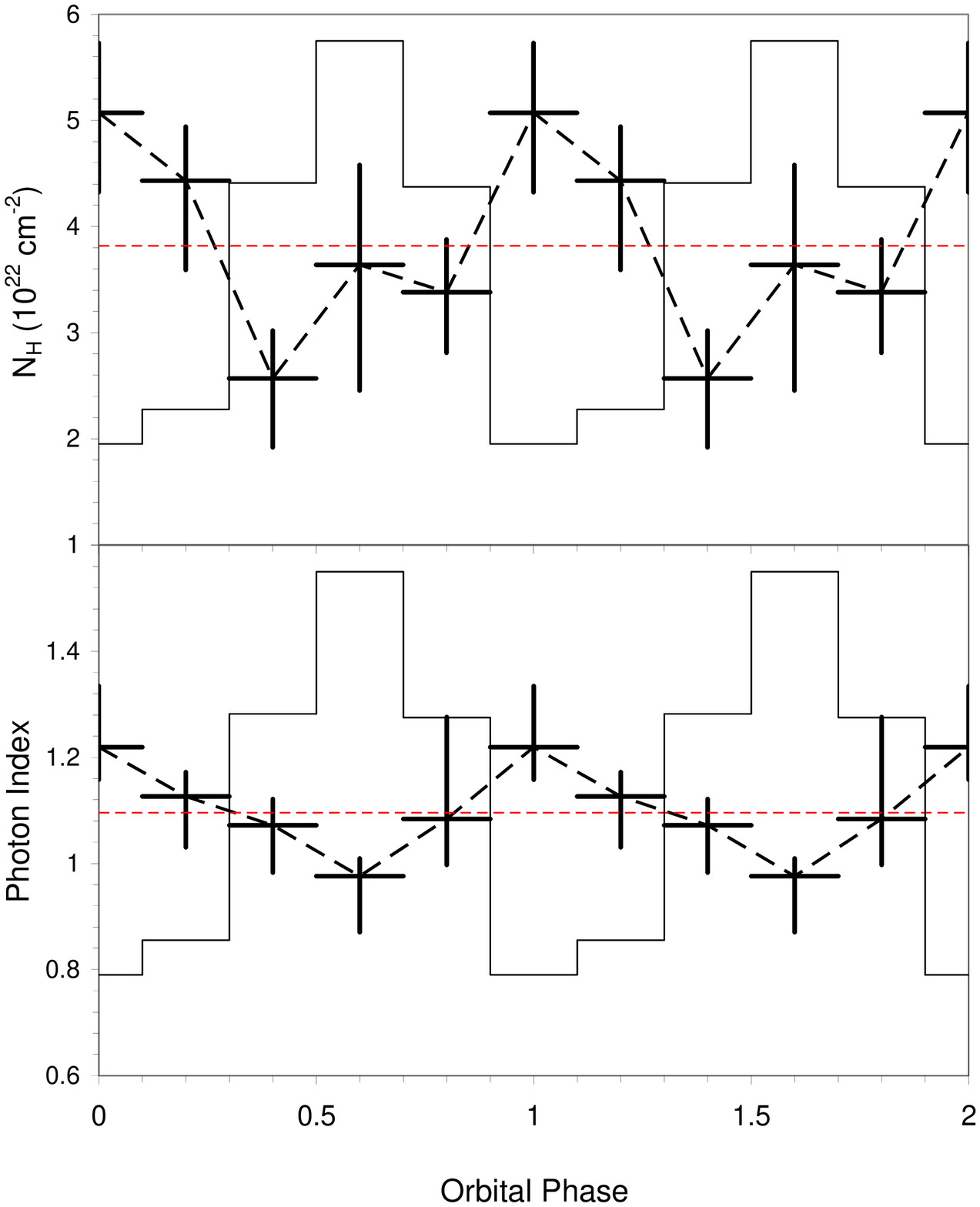}
\caption{Variability of the column density (top) and power law
photon index (bottom) over the orbital period, with error bars
representing 90$\%$ confidence intervals. The histogram represents
the \textit{RXTE} PCA 2 -- 9 keV light curve folded over the
11.599 d orbital period.}\label{specorb1}
\end{center}
\end{figure}

The neutral hydrogen column density and power law photon index
parameters were each plotted against orbital phase, and compared
with the PCA 2 -- 9 keV light curve folded over the same period
(Figure~\ref{specorb1}). The reduced $\chi^2$ values obtained were
2.47 (null hypothesis probability of 4$\%$) and 1.31 (null
hypothesis probability of 26$\%$) for the column density and
photon index respectively, indicating that both parameters are
variable. The linear correlation coefficient values obtained for
the column density and photon index compared to the folded orbital
profiles were -0.78 and -0.84 respectively, indicating strong
anti-correlations between both parameters and the orbital period
folded flux. As with the pulse phase resolved spectral analyses,
we were unable to constrain the equivalent width of the Fe
K$\alpha$ line sufficiently in order to determine whether it is
variable over the orbital period. The orbital phase resolved
spectra are shown in Figure \ref{orbspec} in the Appendix.

\subsection{Super-orbital Variability}

\begin{figure}
\begin{center}
\includegraphics[width=8cm]{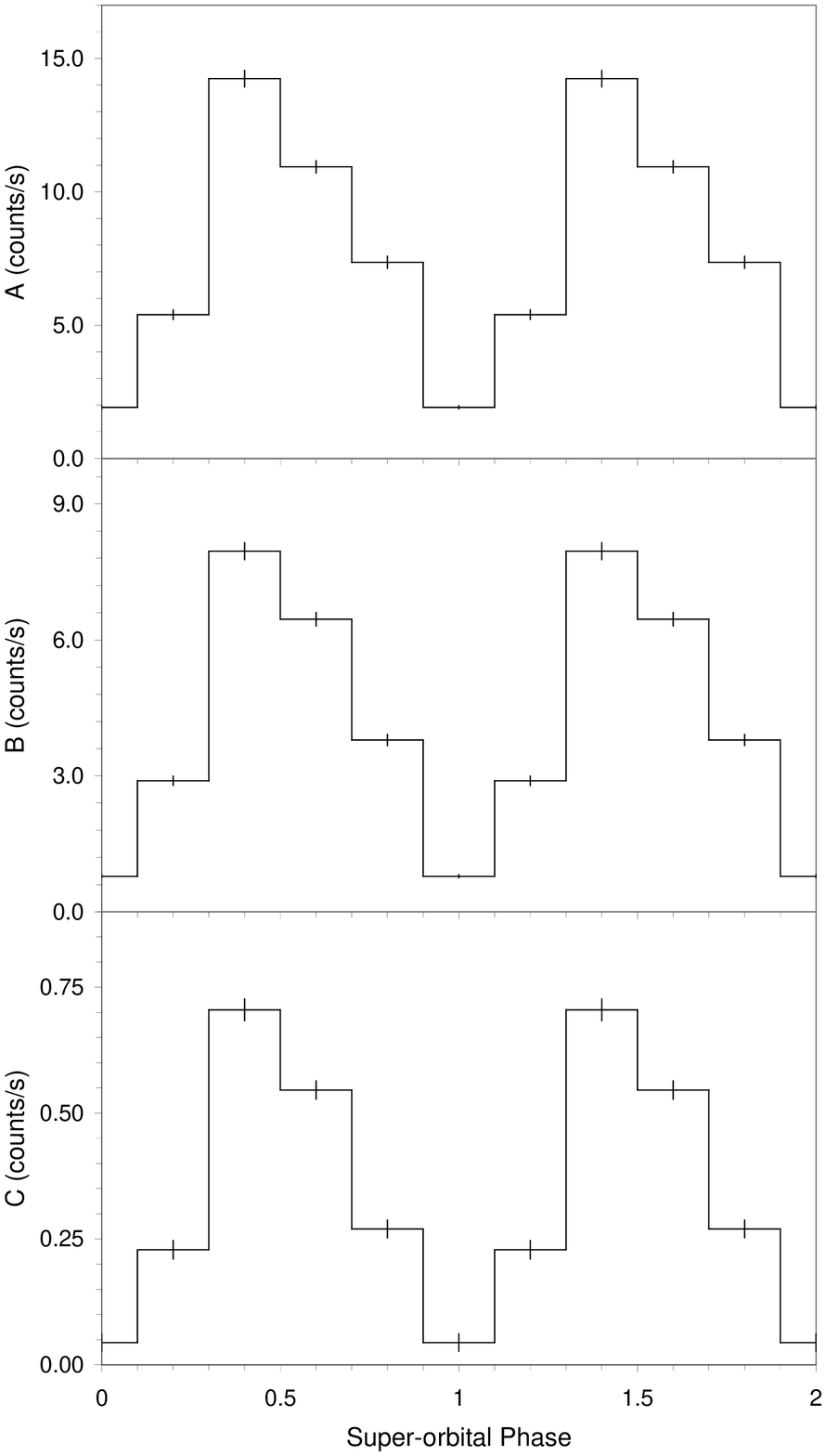}
\caption[PCA Standard2 A (top), B (middle) and C (bottom) light
curves of all observations folded over the 30.75 d super-orbital
period.]{PCA Standard2 A (top), B (middle) and C (bottom) light
curves of all observations folded over the 30.75 d super-orbital
period, with error bars determined from the standard
error.}\label{sorbA}
\end{center}
\end{figure}

As only 2 super-orbital cycles were sampled during the course of
our observations, the value of the super-orbital period could not
be accurately constrained. As with the orbital period analyses,
the value obtained by \citet{wen06} of 30.75 d was used for our
analyses. The A, B and C (Figure~\ref{sorbA}) PCA light curves
covering all observations were folded over 30.75 d, with phase 0
set at the flux minimum at MJD 53488. The super-orbital profile
does not appear to vary with energy.

\begin{figure}
\begin{center}
\includegraphics[width=\columnwidth]{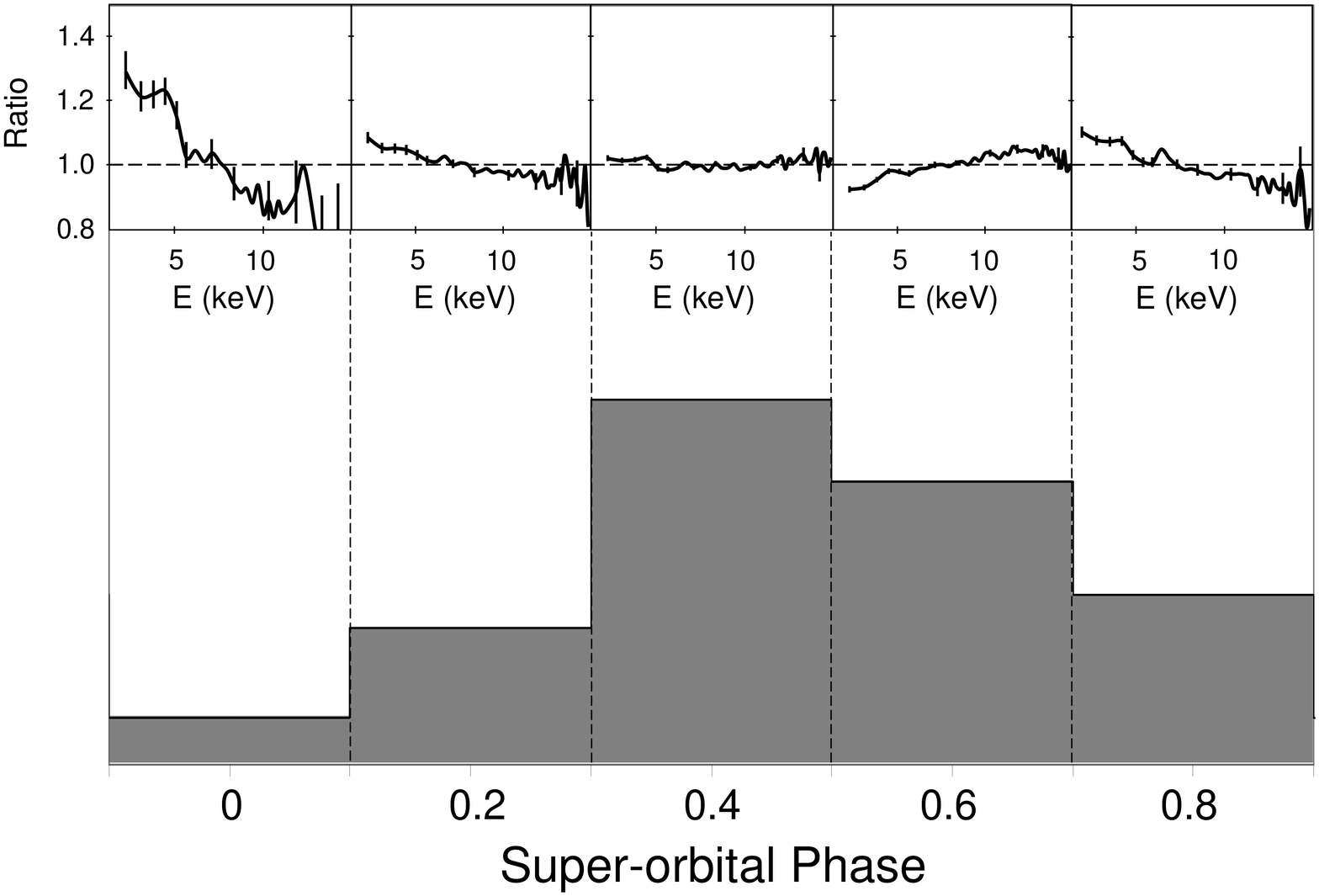}
\caption{Ratios between the super-orbital phase resolved spectra
and the best fit model from the overall spectrum (top inset
figures) plotted against the folded super-orbital profile
(histogram).}\label{sorat}
\end{center}
\end{figure}

Using the methods outlined above for the pulse and orbital period
analyses, we extracted five phase resolved spectra separated by
0.2 phase bins from the PCA data over the 30.75 d super-orbital
period. The ratios between each of the super-orbital phase
resolved spectra and the best-fit model obtained from the overall
spectrum were then plotted. The results are plotted in Figure
\ref{sorat}.

As with the pulse and orbital spectra, the ratio plots indicate
significant spectral variability over the super-orbital period.
Comparison with the results of the simulations presented in Figure
\ref{specvar} indicates likely variability in the slope of the
power law and in the column density parameters, with no evidence
for variability in the high energy cut-off parameters. The same
features around $\sim$5 keV and $\sim$6 keV as observed in the
pulse and orbital plots are present, which we again attribute to
the xenon L-edge and fluorescent iron emission line.

Based on this comparison, the same absorbed power law with high
energy cut-off model was fitted to each of the spectra and the
spectral parameters measured. Exactly the same method was used as
was with the pulse and orbital phase resolved analyses. The
reduced $\chi^2$ statistic obtained from this fitting process was
0.615 with 238 degrees of freedom.

\begin{figure}
\begin{center}
\includegraphics[width=\columnwidth]{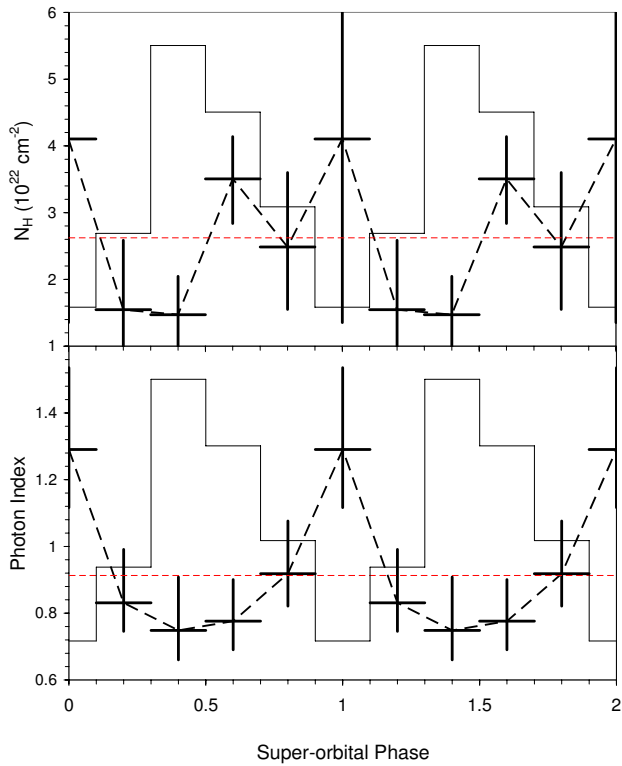}
\caption{Variability of the column density (top) and power law
photon index (bottom) over the super-orbital period, with error
bars representing 90$\%$ confidence intervals. The histogram
represents the \textit{RXTE} PCA 2 -- 9 keV light curve folded
over the best fit 30.75 d super-orbital period.}\label{specsorb1}
\end{center}
\end{figure}

The neutral hydrogen column density and power law photon index
parameters were then plotted against super-orbital phase, and
compared with the PCA 2 -- 9 keV light curve folded over the same
period (Figure~\ref{specsorb1}). The reduced $\chi^2$ values
obtained were 1.88 (null hypothesis probability of 11$\%$) and
1.78 (null hypothesis probability of 13$\%$) for the column
density and photon index respectively, indicating that both
parameters are marginally variable. The linear correlation
coefficient values obtained for the column density and photon
index compared to the folded super-orbital profiles were -0.40 and
-0.84 respectively, indicating a weak and strong anti-correlation
respectively between both parameters and the super-orbital period
folded flux. As with the pulse and orbital analyses, we were
unable to constrain the equivalent width of the Fe K$\alpha$ line
sufficiently in order to determine whether it is variable over the
super-orbital period. The super-orbital phase resolved spectra are
shown in Figure \ref{sorbspec} in the Appendix.

\subsection{Other Variability}

A search for periodic variability between 500 -- 1,500 s was
performed in order to determine whether the flaring episodes which
were seen to occur during the pulse maximum were periodic,
pseudo-periodic or stochastic in nature. The PCA 2 -- 9 keV (A)
light curves for observations 1, 10, 11, 12, 13, 14, 15, 20 and 22
were analysed, as each had good pulse phase coverage and the pulse
modulation could be clearly seen. An \textsc{efsearch} analysis
was performed on each of the light curves in the 500 -- 1,500 s
range with 1 s resolution. The strongest peaks in the $\chi^2$ vs
period plots were found to lie between 1,100 and 1,470 s in all
the data sets except for that pertaining to observation 11, which
showed a significant peak at 710 s. Table~\ref{tab53} lists the
flare recurrence timescales and FWHM uncertainties as measured
from the $\chi^2$ vs period plots.

\begin{figure}
\begin{center}
\includegraphics[width=\columnwidth]{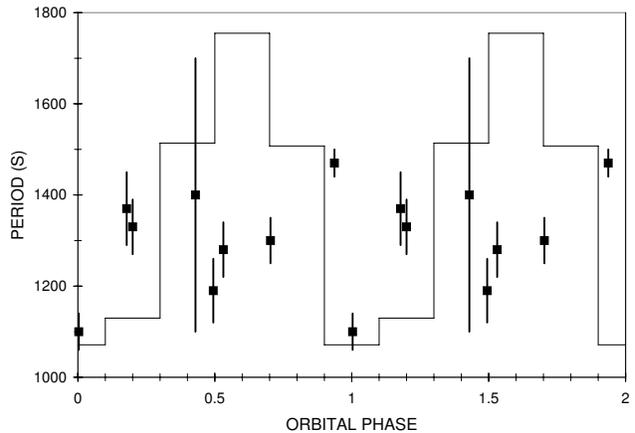}
\caption{Flare recurrence times from Table~\ref{tab53} (squares)
plotted against orbital phase. The histogram represents the
\textit{RXTE} PCA 2 -- 9 keV light curve folded over the 11.599 d
orbital period.}\label{otherorb}
\end{center}
\end{figure}

\begin{figure}
\begin{center}
\includegraphics[width=\columnwidth]{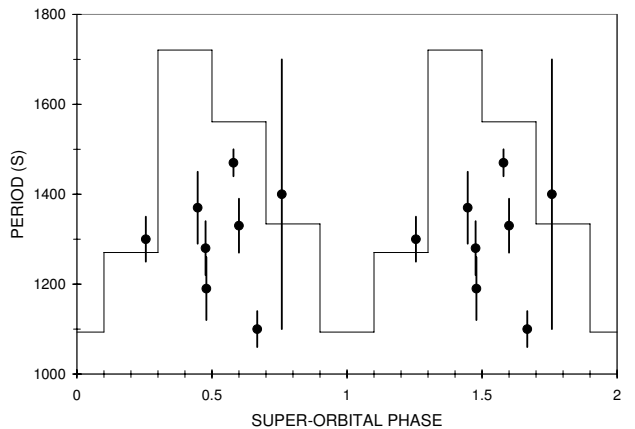}
\caption{Flare recurrence times from Table~\ref{tab53} (circles)
plotted against super-orbital phase. The histogram represents the
\textit{RXTE} PCA 2 -- 9 keV light curve folded over the 30.75 d
super-orbital period.}\label{othersorb}
\end{center}
\end{figure}

The location and width of the strongest peak in each data set
varied considerably with time. In order to determine whether the
flare recurrence times varied smoothly over the orbital or
super-orbital periods, the values in Table~\ref{tab53} were
plotted against orbital (Figure~\ref{otherorb}) and super-orbital
(Figure~\ref{othersorb}) phase, with the 710 s value excluded as
an outlier. No correlation between the flare recurrence times and
the orbital or super-orbital period folded flux profiles was
observed, indicating that the recurrence timescales of the
observed flares do not vary periodically over either the orbital
or super-orbital period.

It is interesting to note, however, that no flares were observed
during the minimum flux of the super-orbital cycle. The flares
were observed in conjunction with the maximum flux in the pulse
cycle (phases $\sim$0.2 -- 0.7), and were completely absent during
the pulse minimum. During the minimum of the super-orbital cycle,
the pulse modulation is absent, and as such no flares were
observed. We interpret this relationship as indicating that the
super-orbital modulation is a result of variable accretion through
some as-yet-unknown mechanism, as the pulse modulation would not
be expected to be observed during phases of low accretion.

\begin{table}
\begin{center}
\caption[Recurrence timescales of the flaring episodes during
pulse maximum]{Recurrence timescales of the flaring episodes
during pulse maximum, with FWHM uncertainties.}\label{tab53}
\begin{tabular}{ccc}
\hline Obs. & Date & Period\\
& (MJD) & (s) \\
\hline
1 & 53506 & 1,330 $\pm$ 60\\
10 & 53533 & 1,300 $\pm$ 50\\
11 & 53535 & 710 $\pm$ 20\\
12 & 53717 & 1,400 $\pm$ 300\\
13 & 53721 & 1,370 $\pm$ 80\\
14 & 53723 & 1,190 $\pm$ 70\\
15 & 53726 & 1,280 $\pm$ 60\\
20 & 53741 & 1,470 $\pm$ 30\\
22 & 53747 & 1,100 $\pm$ 40\\
\hline
\end{tabular}

\end{center}
\end{table}

\section{Discussion}

A total of 22 individual pointed observations were performed with
the \emph{RXTE}, covering 22 pulse cycles, 6 orbital cycles and 2
super-orbital cycles. Significant temporal and spectral
variability was observed from 2S 0114+650 over a range of
timescales. The pulse, orbital and super-orbital modulations were
all clearly present. An additional modulation over timescales of
$\sim$1,300 s was also present, coincident with the maximum of the
pulse period. The overall 3 -- 50 keV spectrum was best fit with
the same absorbed cut-off power law reported by other authors,
albeit without the previously reported Fe K$\alpha$ emission line
at 6.4 keV or the CRSF at 22 keV. The following sub-sections
present a discussion of the major results obtained.

\subsection{Eclipses}

As described in $\S$ 3, the light curve of observation 22
(Figure~\ref{4lc}) exhibits an episode of very low flux which
could be indicative of an eclipse. Comparison with the ASM light
curve folded over the 11.599 d orbital period shows that this
episode of very low-flux does not occur at the expected orbital
minimum but instead just after the maximum. In addition, while the
pulse modulation is not present during observations 4 and 18, it
can be seen clearly in observations 8 and 14, all of which took
place during similar orbital phases as observation 22. Thus, the
low-flux episode in observation 22 is most likely due to a
temporary suspension of accretion, possibly due to density
fluctuations in the stellar wind.

\subsection{Fe K$\alpha$ Emission}

We did not find that the inclusion of an Fe K$\alpha$ Gaussian
emission feature improved the fit significantly in our analysis of
the combined spectrum from all 22 observations. However, during
our phase resolved spectral analyses over the pulse, orbital and
super-orbital periods -- where the 1$\%$ systematic error was
removed and the residuals due to the xenon L-edges were modelled
using an absorption edge, so as to more accurately constrain the
column density (see \S 2) -- significant residuals around
$\sim$6.4 keV were observed in many of the spectra. Addition of an
Fe K$\alpha$ Gaussian emission line was thus required in order to
obtain the best fit. The energy for this line was fixed at 6.4 keV
for the phase resolved analyses, as the energy could not be well
constrained. While the equivalent widths of these lines ranged
between $\sim$60 -- 800 eV, the estimated uncertainties were too
large to determine whether the equivalent widths were correlated
with any of the known periodicities.

\citet{hal00} note that in their observations the equivalent width
of the iron line could only be constrained during the low-state of
the spin period, as the emission feature was most strongly present
during the episodes of low-flux. \citet{mas06} observed the same
phenomenon, concluding that the Fe emission was overwhelmed by the
continuum emission in the high state and was thus only detectable
during the minimum flux periods between pulses. In contrast, the
relatively poor spectral resolution and sensitivity of the
\emph{RXTE} PCA meant that the parameters of the Fe line could not
be adequately constrained in our data. Further observations with
an instrument with greater spectral resolution and sensitivity
than the \emph{RXTE} PCA are required before a firm conclusion can
be drawn as to the nature and variability of the fluorescent iron
emission in this system. Understanding the origin of this emission
could lead to a clearer picture of the geometry and dynamics of
the system, potentially supplying further evidence for the
presence of a super-slow pulsar and providing an answer to the
question of the origin of the super-orbital modulation.

\subsection{Cyclotron Resonance Scattering Features}

The CRSF at $\sim$22 keV reported by \citet{bon05} was not evident
in any of our spectra, supporting the results of \citet{mas06} and
\citet{den06} who also failed to detect this feature.

If the magnetic field of the neutron star in 2S 0114+650 is $\ga$
10$^{12}$ G as suggested by \citet{li99}, we would expect a CRSF
in the spectrum at $\ga$ 9 keV. \citet{hei04} showed there to be a
rough correlation between the energy of the cut-off and the
cyclotron line energy in the spectra of accreting X-ray pulsars
(see their Figure 4). Using a value of $\sim$6 keV for the cut-off
energy as determined from our spectral fits, 2S 0114+650 should
have a CRSF line energy of $\sim$6 keV and therefore a magnetic
field strength of $\sim$10$^{12}$ G.

The scattering cross section has a strong dependence on the angle
to the magnetic field, so that CRSFs provide an excellent
diagnostic of the emission geometry and the physical conditions of
the radiating plasma \citep{har06}. The angle between the photon
direction and the magnetic field vector is thus a determining
factor in the detection of cyclotron features in the spectrum
\citep[e.g.][]{sch07}. The shape and relative depths of the CRSF
first and higher harmonics are thus highly dependent on the
viewing geometry and the geometry of the scattering plasma
\citep{har06}. As a result, variations in the features are
expected with the varying viewing geometry produced by the
rotation of the neutron star. For the case where the photon
direction and magnetic field vector are aligned, no cyclotron
resonance scattering would be expected. In this scenario it may be
possible to place some constraints on the accretion geometry (e.g.
fan beam/cyclinder vs pencil beam/slab geometry), rotation
patterns and accretion rate through the absence of observable CRSF
lines \citep[e.g.][]{sch07}.

Alternatively, the absence of any CRSF features in the X-ray
spectrum could indicate that the magnetic field strength is
outside the observable bandwidth, indicating a field strength of
either $\ll$ 10$^{12}$ G or $\gg$ 10$^{12}$ G. In either of these
scenarios the CRSF would occur at energies too low or too high for
us to observe in our data.

\subsection{The High-Energy Tail}

The detection of a `high energy tail' in the spectrum of 2S
0114+650 from analysis of $\emph{INTEGRAL}$ IBIS-ISGRI data was
reported by \citet{den06}, which required a combination of thermal
bremsstrahlung and power law models to describe the spectrum at
high energies. Their findings are thus inconsistent with those of
\citet{hal00}, \citet{mas06}, \citet{muk06} and with our own
results.

In order to test the validity of their model, we attempted to fit
a thermal bremsstrahlung + power law model to our combined PCA +
HEXTE spectra in the 3 -- 50 keV range. While this model provided
an acceptable fit, significant residuals were seen at $\sim$5 keV
indicating that the model does not accurately describe the
spectrum. Due to the poor statistics in the HEXTE data above 50
keV we could not confirm the existence of a tail in the spectrum
$>$ 70 keV. However, the poor residuals at low energies obtained
with the combined thermal bremsstrahlung/power law model raises
questions about the presence of a high energy tail in the spectrum
from this system. It should be noted that \citet{den06} also found
an acceptable fit to the spectrum using a single power law model
without the inclusion of a high energy exponential cut-off.

Additionally, in other binary systems the appearance of a high
energy tail has been linked to the onset of compact jets. Radio
observations of 2S 0114+650 performed in the 240 -- 1400 MHz range
with the Giant Metrewave Radio Telescope found no evidence for any
radio emission, thus arguing against the presence of jets in this
system \citep{far07}.

\subsection{The $\sim$2.7 hr Pulse Period}

From our analysis of the PCA timing data, we have obtained a new
value for the pulse period at 2.65 $\pm$ 0.003 h. This value is
inconsistent with the value reported by \citet{bon05}, indicating
that the neutron star is continuing to spin-up over time
\citep[see][]{hal00,soo06}. Separate analysis of the two
observation runs showed no significant difference between the
values of the pulse period during the six months separating the
runs, providing an upper limit for $\dot{P}_{spin}$/$P_{spin}$
during MJD 53505 -- 53747 of $\la$ $-$ 4.6 $\times$ 10$^{-3}$
yr$^{-1}$. Our results are thus consistent with the value of
$\sim-$2 $\times$ 10$^{-3}$ yr$^{-1}$ obtained by \citet{hal00}.

Our results also indicate that the model proposed by \citet{kon06}
-- which ascribed the $\sim$2 h pulse period to tidally-induced
pulsations of the B-supergiant companion -- is inconsistent with
the observed properties of this source. They showed that tidal
oscillations of the companion star's surface could occur at both
the $\sim$2 h and $\sim$30 d timescales. However, their model
predicts that the $\sim$2.7 h period should vary significantly on
short times-scales between $\sim$1.6 -- 3.4 h (depending on the
system parameters). In addition, the link between tidal
oscillations of the donor star and periodic mass ejection episodes
is still unclear \citep{kon06}. Our observations thus indicate
that this model does not adequately explain the mechanism behind
the pulse period, as the stability of this period has been
demonstrated over timescales of days to months.

The pulse profile shown in Figure \ref{pulseA} is approximately
sinusoidal with features around phase 0.2 and 0.85, with the
latter possibly arising from the edge of the emission region at
the second pole \citep[e.g.][]{wan81}. The profile of the
modulation does not appear to vary significantly with energy,
implying an absence of cyclotron scattering \citep[as any
cyclotron features in the spectrum would be expected to vary with
the rotation of the magnetic field; e.g.][]{har06}, supporting the
results of our spectral fitting. In addition, while no clear-cut
dependence of the pulse profile on the pulse period is apparent
for X-ray pulsars, \citet{wan81} noted that many slow pulsars
(P$_{spin}$ $>$ 100 s) show relatively smooth and sinusoidal
profiles. Our results thus continue to support a pulsar nature for
the compact object in 2S 0114+650.

The approximately sinusoidal profile and the lack of change with
energy is consistent with the profiles of other X-ray pulsars with
luminosities below $\sim$10$^{36}$ erg s$^{-1}$ \citep*{whi83}.
Previous attempts to constrain the distance to 2S 0114+650 have
yielded a range of values \citep[1.8 $\pm$ 0.2 kpc, $\sim$2.5 kpc,
$\sim$3.8 kpc, $\sim$3
kpc;][respectively]{aab83,cra85,che98,kon03}, with the most
thorough attempt by \citet{rei96} providing a poorly constrained
distance of 7.0 $\pm$ 3.6 kpc. Taking 10$^{36}$ erg s$^{-1}$ as
the luminosity upper-limit in the 0.5 -- 60 keV band from
\citet{whi83} we can provide further constraints on the distance.
Assuming a lower limit of 3.4 kpc from \citet{rei96}, we derive a
new distance estimate of 4.5 $\pm$ 1.5 kpc.

Our analysis of the phase resolved spectra showed marginal
evidence for variability in the power law photon index over the
pulse period, anti-correlated with the flux. A similar
anti-correlation between the photon index and the pulse flux has
been observed in Her X-1 \citep[e.g.][]{soo90}. The observed
similarities between the spectral variability of 2S 0114+650 and
Her X-1 over the pulse periods could thus strengthen the case that
the $\sim$2.7 h period is indicative of the neutron star spin
period. However, the uncertainties for the photon index values are
too large to draw a firm conclusion either way. No significant
correlated variability of the neutral hydrogen column density over
the pulse period could be determined.

Using the derived values of the wind velocity \citep[1,200 km
s$^{-1}$;][]{cor87}, orbital tangential velocity ($\sim$271 km
s$^{-1}$ assuming an almost-circular orbit), orbital angular
velocity (6.3 $\times$ 10$^{-6}$ rad s$^{-1}$), and typical values
for the other parameters \citep{ikh06}, we calculated the
$\dot{M}$ in order to achieve the observed average spin-up rate.
The results of these calculations yielded values of $\sim$9
$\times$ 10$^{-9}$ M$_\odot$ yr$^{-1}$ for a neutron star and
$\sim$2 $\times$ 10$^{-3}$ M$_\odot$ yr$^{-1}$ for a white dwarf.
The derived mass accretion rate for a neutron star is
approximately two orders of magnitude less than the mass loss rate
for B1 Ia supergiants derived by \citet{bar77} from infrared
observations and, more recently, by \citet{cro06} from optical
observations.

The mass accretion rate would be expected to be much less than the
mass loss rate of the donor star due to inefficient capture of
material from the stellar wind. Our derived mass accretion rate is
thus consistent with the expected mass loss rate of the
supergiant. In contrast, the derived mass accretion rate for a
white dwarf is three orders of magnitude greater than the expected
supergiant mass loss rate, clearly excluding a white dwarf as the
compact object in this system. In light of this, the evolution of
the pulse period \citep{soo06}, and the spectral similarities
shared with other binary X-ray pulsars, we conclude that the
$\sim$2.7 h period must indeed be the spin-period of a neutron
star.

The recent reports by \citet{mat06} and \citet{del06} of two other
candidate pulsars with hour-long spin-periods ($\sim$5 h for 4U
1954+319 and $\sim$6.7 h for 1E 161348-5055.1 respectively) -- if
confirmed -- would appear to indicate that super-slow rotating
neutron stars are more prevalent than originally suspected.
\citet{ikh06} has shown that the assumption of a super-critical
initial magnetic field strength (i.e. $B \gg 10^{13}$ G) is not
necessary if the propeller phase consisted of both supersonic and
subsonic propeller phases. Rapid spin-down would occur over very
short timescales due to interactions between the accreting
material and the neutron star magnetic field. In this setting,
even spherical accretion from a stellar wind would result in
significant angular momentum transfer due to the orbital motion of
the neutron star \citep{ikh06}, leading to spin-down timescales
consistent with the projected lifetime of the donor star.

\citet{ikh06} showed that in the Davies-Fabian-Pringle
\citep{dfp79} scenario a mass accretion rate of $\sim$10$^{14}$ g
s$^{-1}$ is required to spin 2S 0114+650 down to a period of
$\sim$10$^4$ s. This derived accretion rate is a factor of 30
smaller than the current mass transfer rate inferred from the
X-ray luminosity of the source. However, the evolutionary tracks
of massive stars predict a main sequence O9.5 V progenitor star
for a B1 Ia supergiant in a previous epoch \citep{mey94}, with a
mass outflow rate of 30 -- 40 times smaller than the donor star in
2S 0114+650 inferred from optical observations. Hence, the mass
accretion rate in a previous (spin-down) epoch would be
$\sim$10$^{14}$ g s$^{-1}$, consistent with the value required by
the Davies-Fabian-Pringle scenario.

The problem common to both the \citet{li99} and \citet{ikh06}
scenarios is that the equilibrium period for 2S 0114+650 is
predicted to be $\la$ 26 min, which the spin-period will rapidly
approach on timescales of $<$ 100 yr for disc accretion and $<$
1,000 yr for stellar wind accretion \citep{ikh06}. The probability
of observing the pulsar in 2S 0114+650 during the long-period
phase is thus almost negligible, indicating either 2S 0114+650 is
a young accretor (and we have beaten all the odds to observe it in
this stage) or our current understanding of pulsar evolution is
deficient. If the pulse period continues to decrease at the
current average rate of $\sim-$2 $\times$ 10$^{-3}$ yr$^{-1}$
\citep{hal00}, the pulsar will reach its equilibrium period in
$\la$ 400 yr. This timescale is thus consistent with an infant
X-ray pulsar spinning-up through stellar wind accretion.

\subsection{The $\sim$11.6 d Orbital Period}

\begin{figure}
\begin{center}
\includegraphics[width=\columnwidth]{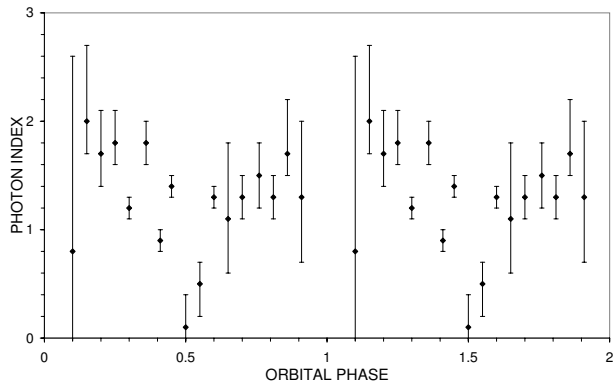}
\caption{Variability of the photon index over orbital phase of EXO
1722-363. Phase 0.5 corresponds to the maximum of the orbital
period, with phases 0.9 -- 1.1 corresponding to the phase of
eclipse \citep[data taken from][]{tho07}.}\label{exopi}
\end{center}
\end{figure}

The observed variability of the neutral hydrogen column density
over the orbital period indicates that variable absorption by the
dense stellar wind throughout the orbit is the mechanism behind
the orbital modulation, as suggested by \citet{hal00} and
\citet{gru07}. Similar variability in the column density has been
observed in the HMXB pulsars Vela X-1 and EXO 1722-363
\citep{nag91,tho07}. The latter system in particular provides a
good comparison, as both the column density and the power law
photon index were observed to vary over the orbital period,
anti-correlated with flux \citep[see Figure \ref{exopi} for a plot
of the photon index over orbital phase from the data presented
in][]{tho07}. EXO 1722-363 is a highly obscured binary system that
exhibits regular eclipses. The observed increase in the column
density and photon index during eclipse are thus a result of
absorption effects.

It is not possible to determine from our results whether accretion
in an unperturbed stellar wind is sufficient to account for the
observed changes in the column density. Thus, we cannot determine
whether structures such as accretion wakes \citep[similar to those
detected in Vela X-1 and EXO 1722-363;][]{hab90,tho07} are
required to account for the observed absorption. However, the
shape of the column density variations with orbital phase is
consistent with an undisturbed wind, and not like the highly
disturbed wind models obtained from simulations \citep[see Figures
2 and 8 of][]{blo91}. Further observations and modelling are
required before the geometry of the wind can be constrained.

While our results do not exclude periodic eclipses of the neutron
star by the supergiant companion, we found no strong evidence for
eclipses in the PCA light curves. We thus conclude that 2S
0114+650 does not exhibit the properties of an eclipsing binary
system as suggested by \citet{cor99}. However, the similarities
between the spectral variability in 2S 0114+650 and EXO 1722-363
hint at similarities in the orbital properties. We thus ascribe
the observed spectral variability to 2S 0114+650 passing behind a
heavily absorbing region close to the base of the wind of the
donor star.

\subsection{The $\sim$30.7 d Super-orbital Period}

Her X-1 is the prototypical system in which the warped, precessing
disc scenario has been shown to explain the observed super-orbital
modulation very well. The column density in this system has been
observed to vary out of phase with the super-orbital cycle, due to
variable obscuration of the central source by the precessing disc
warp \citep{nai03}. The lack of variability of the column density
over the super-orbital period in 2S 0114+650 thus indicates that
this modulation is not due to local changes in the column density,
and therefore precludes variable absorption by a warped,
precessing accretion disc as the cause of the super-orbital
modulation.

The lack of significant Fe K$\alpha$ emission precludes variable
reflection by a precessing disc, as strong iron emission due to
reprocessing of X-rays in a disc would be expected to be present
and variable in this scenario \citep[e.g.][]{geo91}. In addition,
the apparent increase in the power law photon index during the
super-orbital minimum is in exact opposition to the behaviour of
two of the other well studied super-orbital systems -- LMC X-4 and
Her X-1 \citep{nai03} -- both of which exhibit photon index
variability modulated over the super-orbital period. In both of
these systems the super-orbital modulation is attributed to the
warped precessing disc model, once again showing that the 30.75 d
period in 2S 0114+650 does not fit the standard model.

If the super-orbital modulation was tied to accretion onto another
compact object (i.e. a triple system), we would expect to see
variability of the column density and possibly the photon index in
a similar fashion to the orbital mechanism discussed above. While
modulation of the eccentricity of the 11.6 d orbit by a third
massive body would not lead to variability in the column density,
this model would yield a long-period of $\gg$ 30.75 d.

Similarly, if precession of the neutron star-spin axis were the
cause of the super-orbital modulation, the precession period would
be inversely proportional to the spin period \citep{was03}. We
would thus expect to see the super-orbital period vary,
anti-correlated with the spin-up of the neutron star. The apparent
long-term stability of this period as shown by \citet{soo06}
indicates that this latter scenario is unlikely. In addition, the
expected precession period for a neutron star spinning with a
$\sim$2.7 hr period is $\gg$ 30.75 d.

The periodic episodes of enhanced mass accretion in the
tidally-driven oscillation model proposed by \citet{kon06} would
not be expected to cause column density variability. \citet{hud78}
analysed 1,250 photographic plates taken of 2S 0114+650 between
1928 and 1977 at optical wavelengths and could not discern any
significant variability in the optical magnitude greater than 0.3
mag over any timescale. Pulsations in the supergiant donor star
should result in significant variable optical luminosity thus
making this scenario unlikely. However, it should be noted that no
long-term optical monitoring has been performed since 1999 when
the super-orbital period was first detectable in the ASM data,
preventing us from ruling out this model as an explanation for the
super-orbital variability.

Another possibility is that the neutron star undergoes periods of
enhanced accretion over the super-orbital timescale due to the
presence of another donor star in an eccentric orbit of $\sim$30
d. As the second donor star approaches periastron passage,
accretion onto the neutron star would commence through Roche lobe
overflow. The increase in flux over the super-orbital period would
thus be due to phase dependent episodes of Roche lobe overflow,
which would be associated with the formation of a transient disc
during very specific super-orbital phases. This would result in an
increase in the accretion rate and therefore an increase in flux,
but would not necessarily result in variability of the observed
spectral parameters. Previous optical observations have identified
only one candidate for the optical counterpart within the X-ray
error circle. If the super-orbital modulation is due to the
presence of a second donor star in a triple system, the third star
would have to be low-mass with very low luminosity.

\subsection{Other Variability}

Pseudo-periodic flaring episodes were observed around $\sim$1,300
s at the maximum pulse phase, with the recurrence timescale of
these flares seen to vary significantly over time. No evidence for
a correlation between the flare recurrence timescales and either
the orbital or super-orbital period flux was found. When taken
into context with the 894, 850 and $\sim$2,000 s periods
previously reported by \citet{kon83}, \citet{yam90} and
\citet{apa91} respectively and the X-ray flickering reported by
\citet{mas06}, we conclude that the observed $\sim$1,300 s flares
are most likely due to an aperiodic process with a stochastic
recurrence timescale. Possible explanations include ``clumping''
of the material in the accretion stream due to turbulent
processes, or possibly feedback between the radiation pressure
from the emission region and the accreting material. Both
explanations imply a correlation between
the flare recurrence timescales and the mass accretion rate.\\

\section{Conclusions}

We have performed an extensive study of the X-ray variability of
2S 0114+650 with the $\emph{RXTE}$ satellite. The good spectral
coverage provided by the PCA and HEXTE instruments have allowed us
to undertake the first in-depth wide-band study of the
super-orbital modulation. The overall X-ray spectrum in the 3 --
50 keV range was found to be well described by the standard
absorbed power law model with a high energy exponential cut-off,
with spectral parameters consistent with those reported by other
authors from other instruments and epochs.

Marginal evidence for variability in the power law photon index
over the pulse period was identified, which could be attributed to
the varying viewing geometry of the accretion region with the spin
of the neutron star. Similar variability has been observed from
other X-ray pulsars such as Her X-1, supporting the conclusion
that 2S 0114+650 contains a super-slow X-ray pulsar. The
near-sinusoidal pulse profile and the lack of change with energy
puts 2S 0114+650 in the class of low luminosity ($\la$ 10$^{36}$
erg s$^{-1}$) X-ray pulsars, allowing us to provide a new distance
estimate of 4.5 $\pm$ 1.5 kpc.

Variability of the neutral hydrogen column density over the
orbital period was observed, indicating that the line-of-sight
motion of the neutron star through the dense circumstellar
environment is the mechanism behind the orbital modulation.
Additional variability in the power law photon index
anti-correlated with flux over the orbital period was identified,
which we speculate is due to absorption effects as the neutron
star passes behind a heavily absorbing region near the base of the
supergiant companion's wind.

In contrast, no significant variability of the column density was
observed over the super-orbital period. While super-orbital
periods are commonly attributed to variable obscuration by a
precessing warp in an accretion disc, the absence of super-orbital
column density variability indicates that this is not the
mechanism behind the super-orbital modulation. Similarly, the
absence of significant iron emission or reflection components in
the spectrum argues against variable reflection by a precessing
warped disc. While the slope of the spectrum was relatively stable
throughout the super-orbital cycle, a significant increase in the
photon index was observed during the minimum phase. We conclude
that the super-orbital period is tied to variability in the mass
accretion rate due to some as yet unidentified mechanism.

\section*{Acknowledgments}

SAF acknowledges the financial support provided by the UNSW@ADFA
UCPRS scholarship scheme. We thank the entire \emph{RXTE} team for
acquiring the X-ray data and advice on the data analysis. We thank
Natalie Webb and Didier Barret for their many useful discussions
and the anonymous referee for their excellent comments which
significantly improved this paper.

\appendix

\section[]{Individual Light Curves}

The individual light curves for the 2 -- 9 keV (A) energy band
from each of the 22 observations are included in this Appendix.

\begin{figure*}
\begin{center}
\includegraphics[scale=0.9]{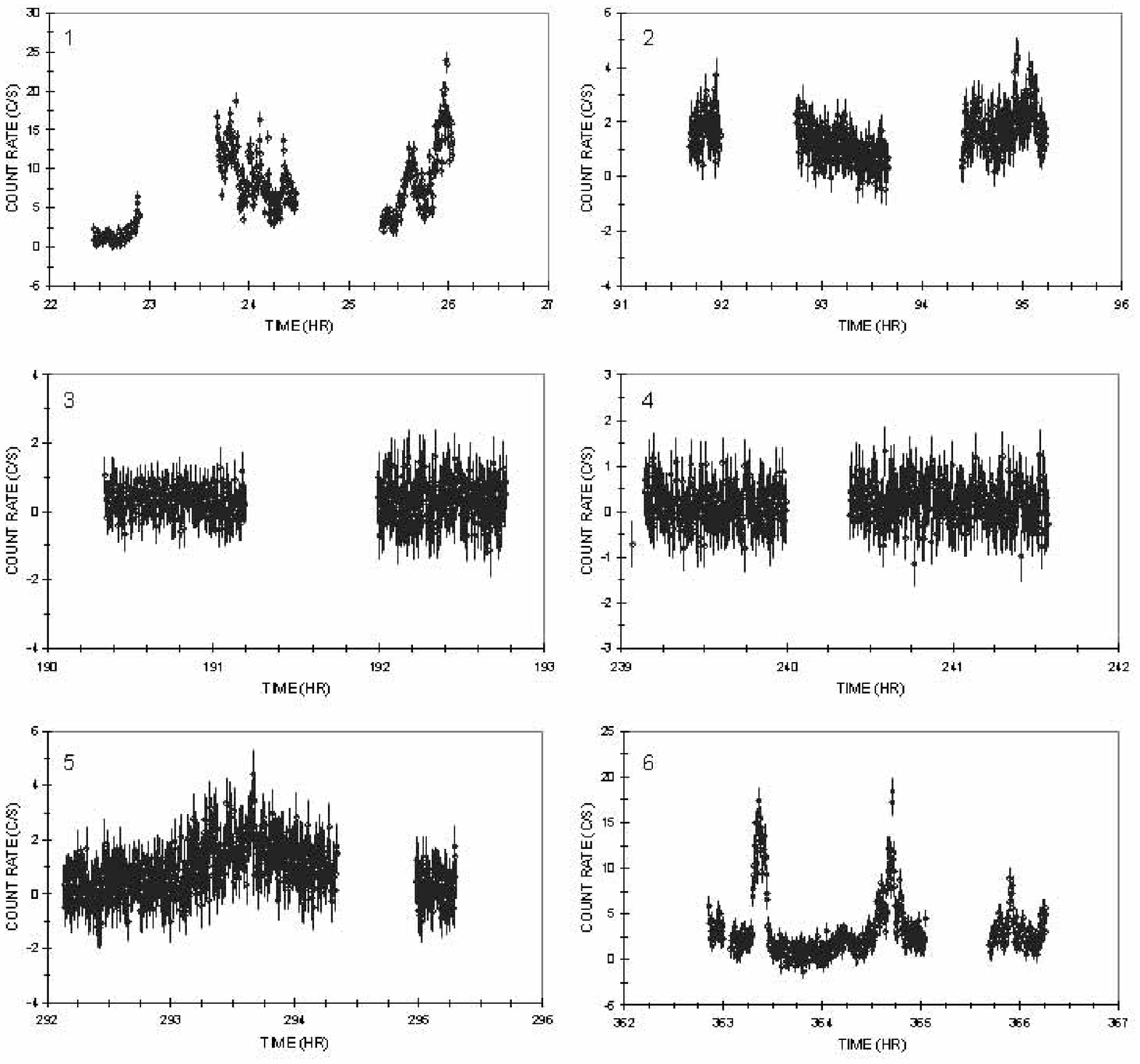}
\caption{The 2 -- 9 keV PCA light curves for observations 1 to 6,
with the time axes showing hours elapsed since MJD
53505.}\label{1lc}
\end{center}
\end{figure*}

\begin{figure*}
\begin{center}
\includegraphics[scale=0.9]{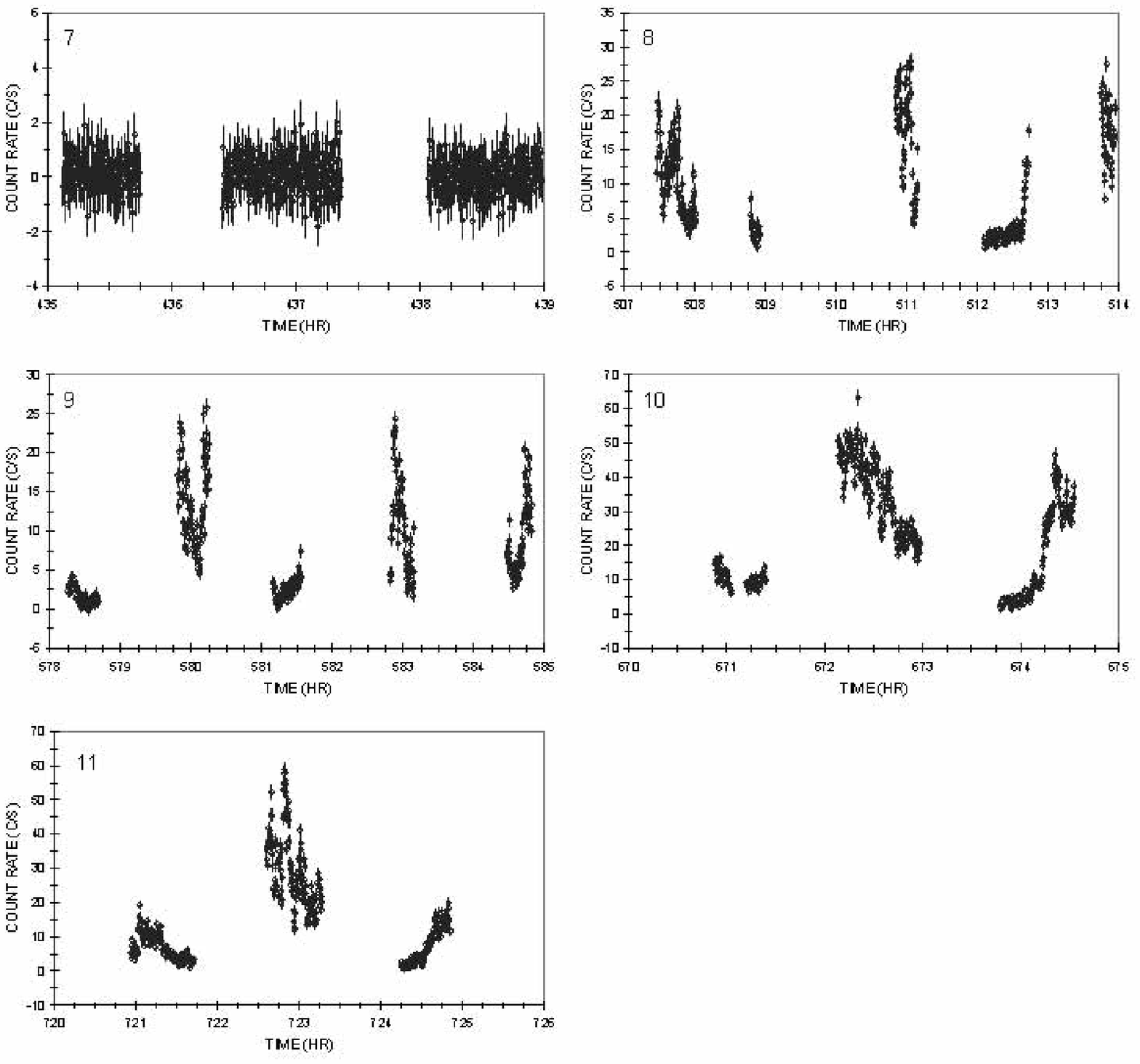}
\caption{The 2 -- 9 keV PCA light curves for observations 7 to 11,
with the time axes showing hours elapsed since MJD
53505.}\label{2lc}
\end{center}
\end{figure*}

\begin{figure*}
\begin{center}
\includegraphics[scale=0.9]{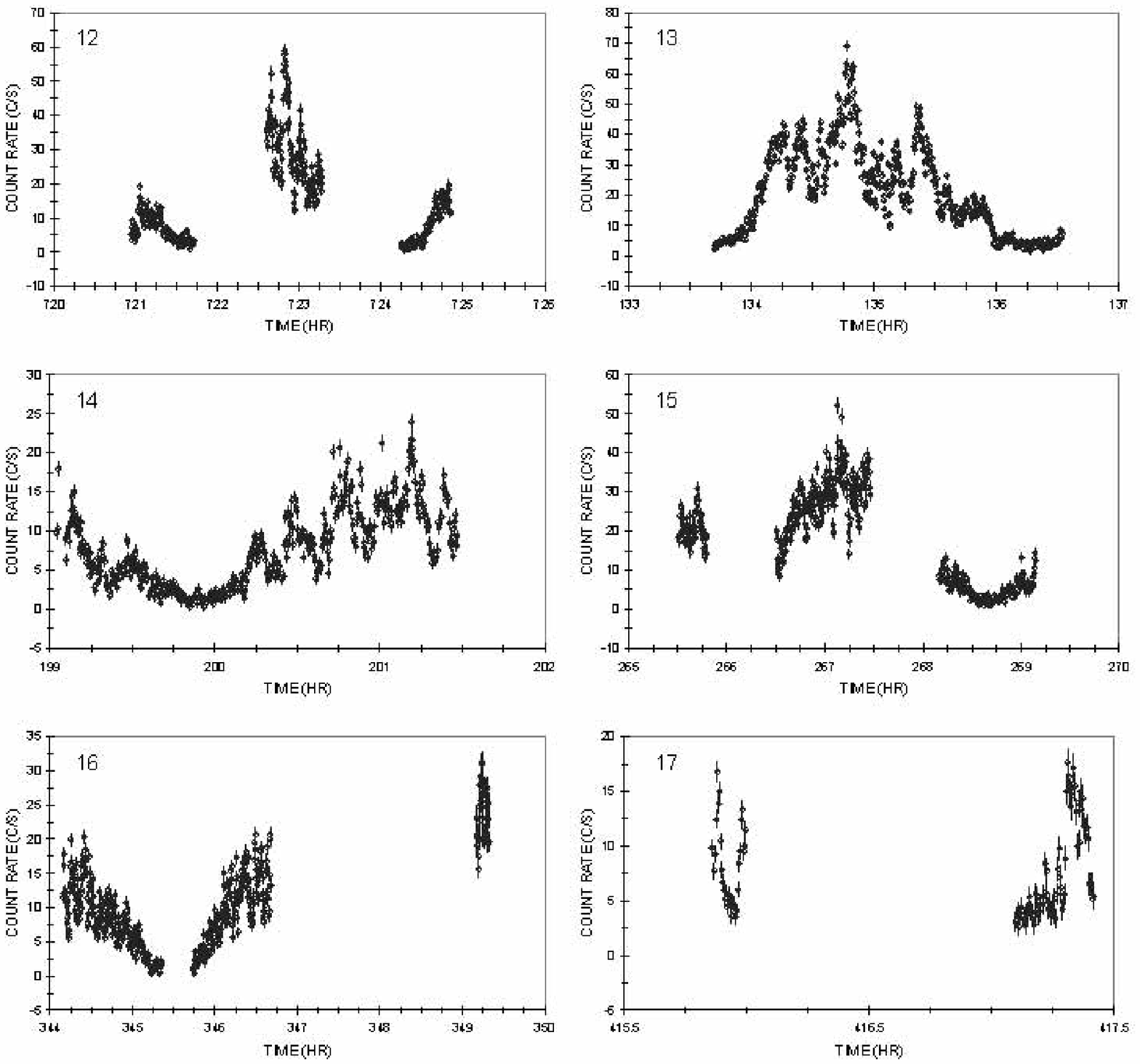}
\caption{The 2 -- 9 keV PCA light curves for observations 12 to
17, with the time axes showing hours elapsed since MJD
53715.}\label{3lc}
\end{center}
\end{figure*}

\begin{figure*}
\begin{center}
\includegraphics[scale=0.9]{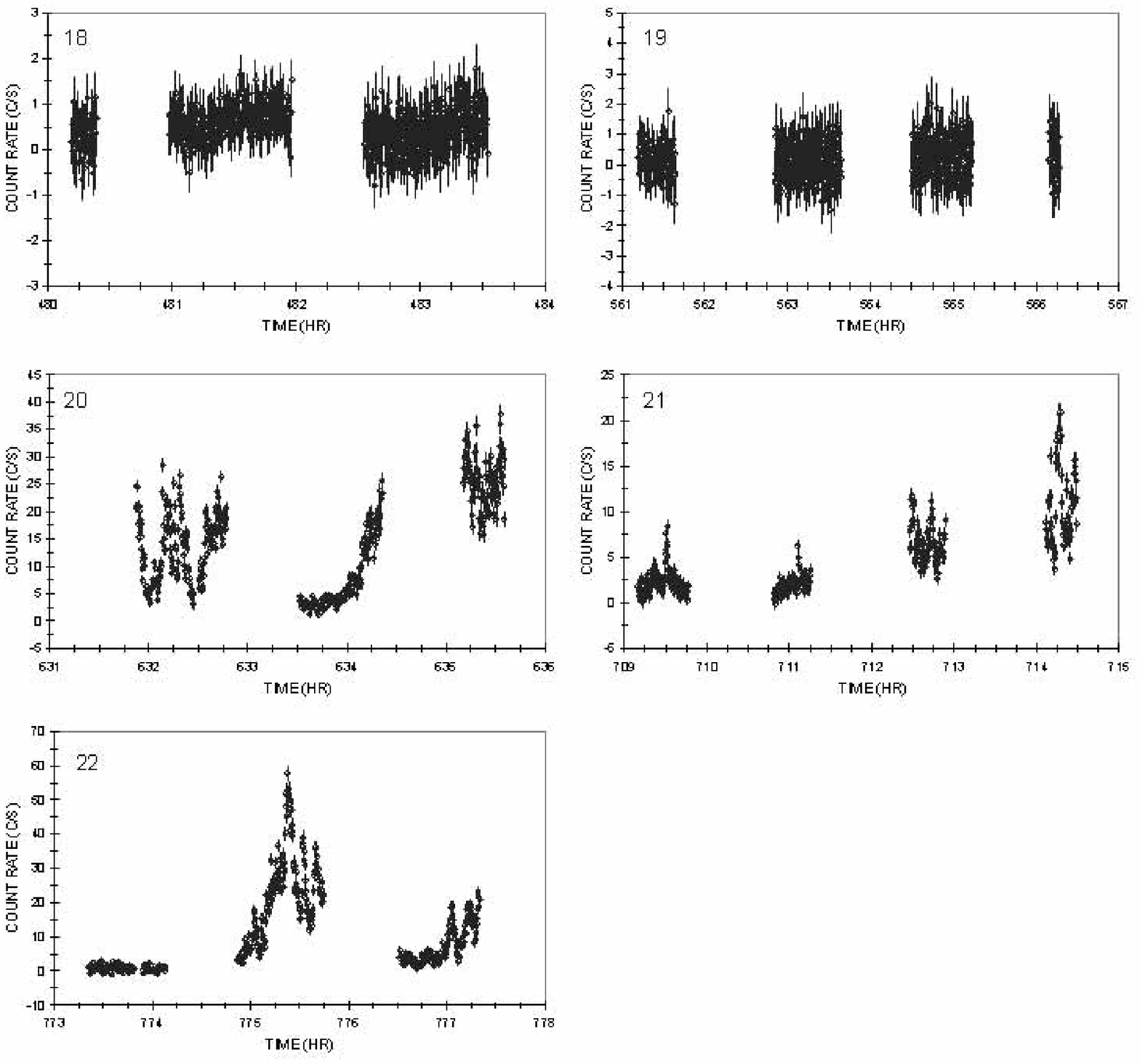}
\caption{The 2 -- 9 keV PCA light curves for observations 17 to
22, with the time axes showing hours elapsed since MJD
53715.}\label{4lc}
\end{center}
\end{figure*}

\section[]{Phase Resolved Spectra}

The spectra from the pulse, orbital and super-orbital phase
resolved analyses are included in this Appendix.

\clearpage

\begin{figure*}
\begin{center}
\includegraphics[width=14cm]{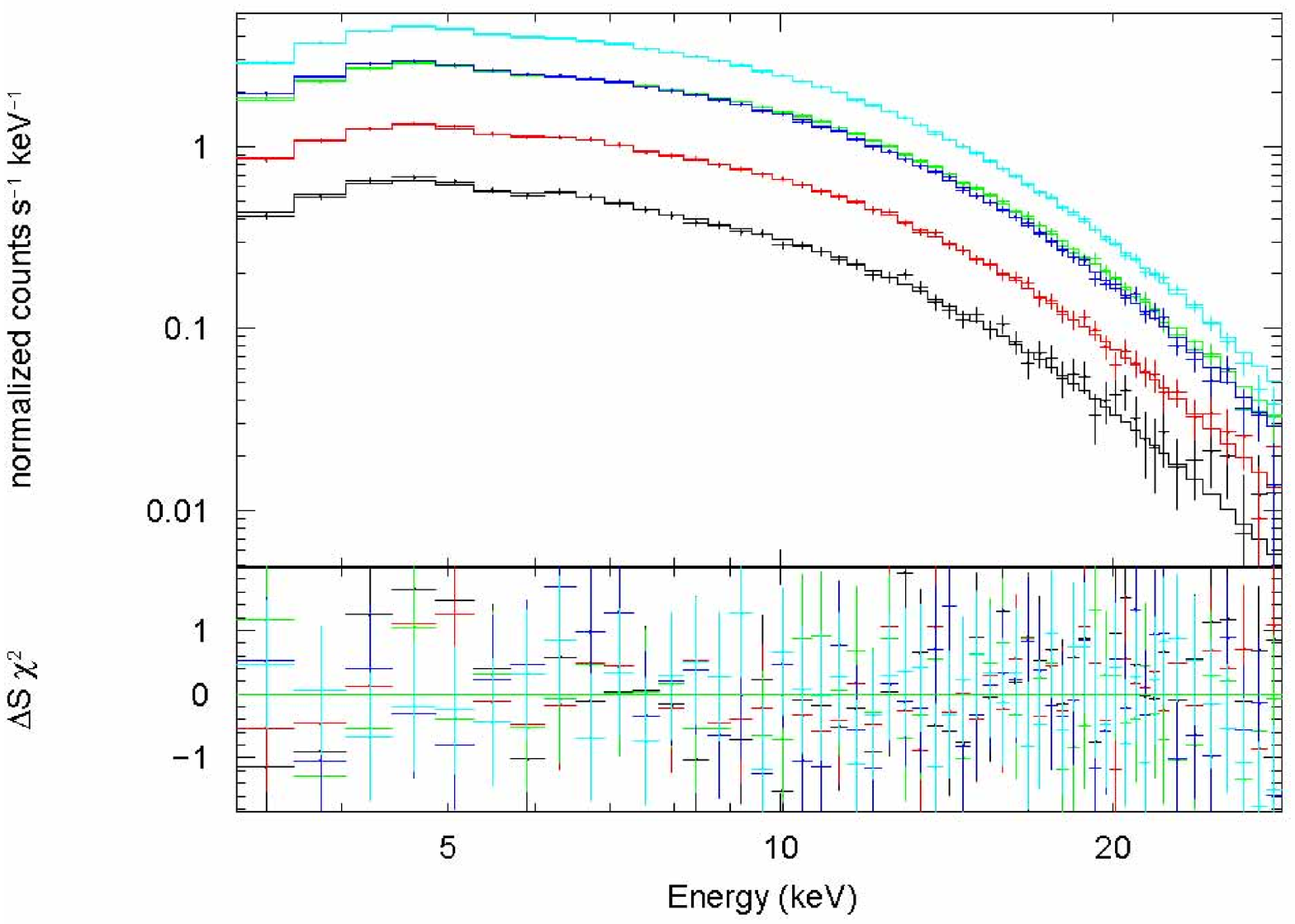}
\caption{PCA 3 -- 30 keV spectra from the pulse phase resolved
spectral analyses for phases 0 to 0.4, ordered from bottom to top
at 5 keV: 0, 0.1, 0.2, 0.3, 0.4.}\label{pulsespec1}
\end{center}
\end{figure*}

\begin{figure*}
\begin{center}
\includegraphics[width=14cm]{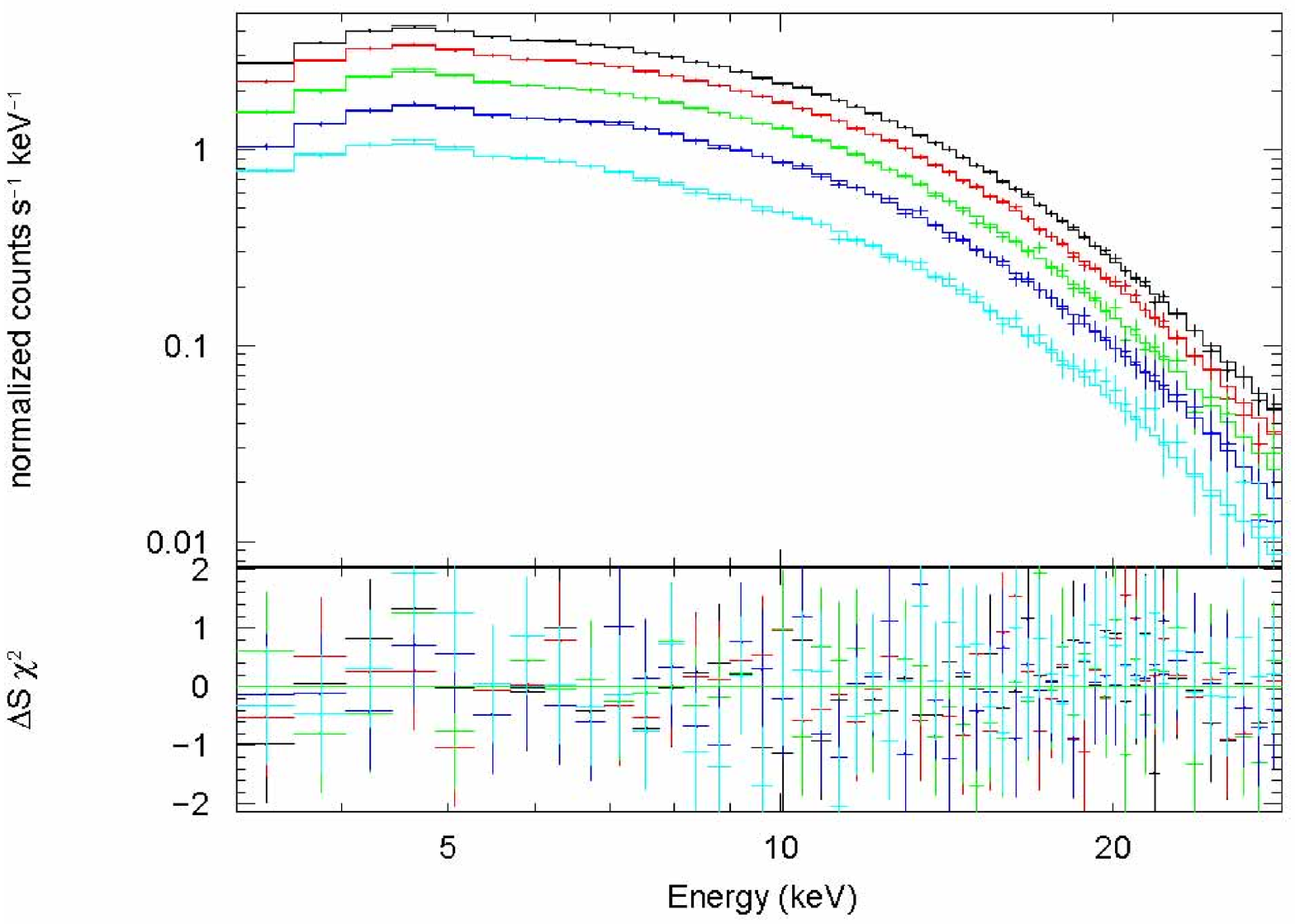}
\caption{PCA 3 -- 30 keV spectra from the pulse phase resolved
spectral analyses for phases 0.5 to 0.9, ordered from bottom to
top at 5 keV: 0.9, 0.8, 0.7, 0.6, 0.5.}\label{pulsespec2}
\end{center}
\end{figure*}

\begin{figure*}
\begin{center}
\includegraphics[width=14cm]{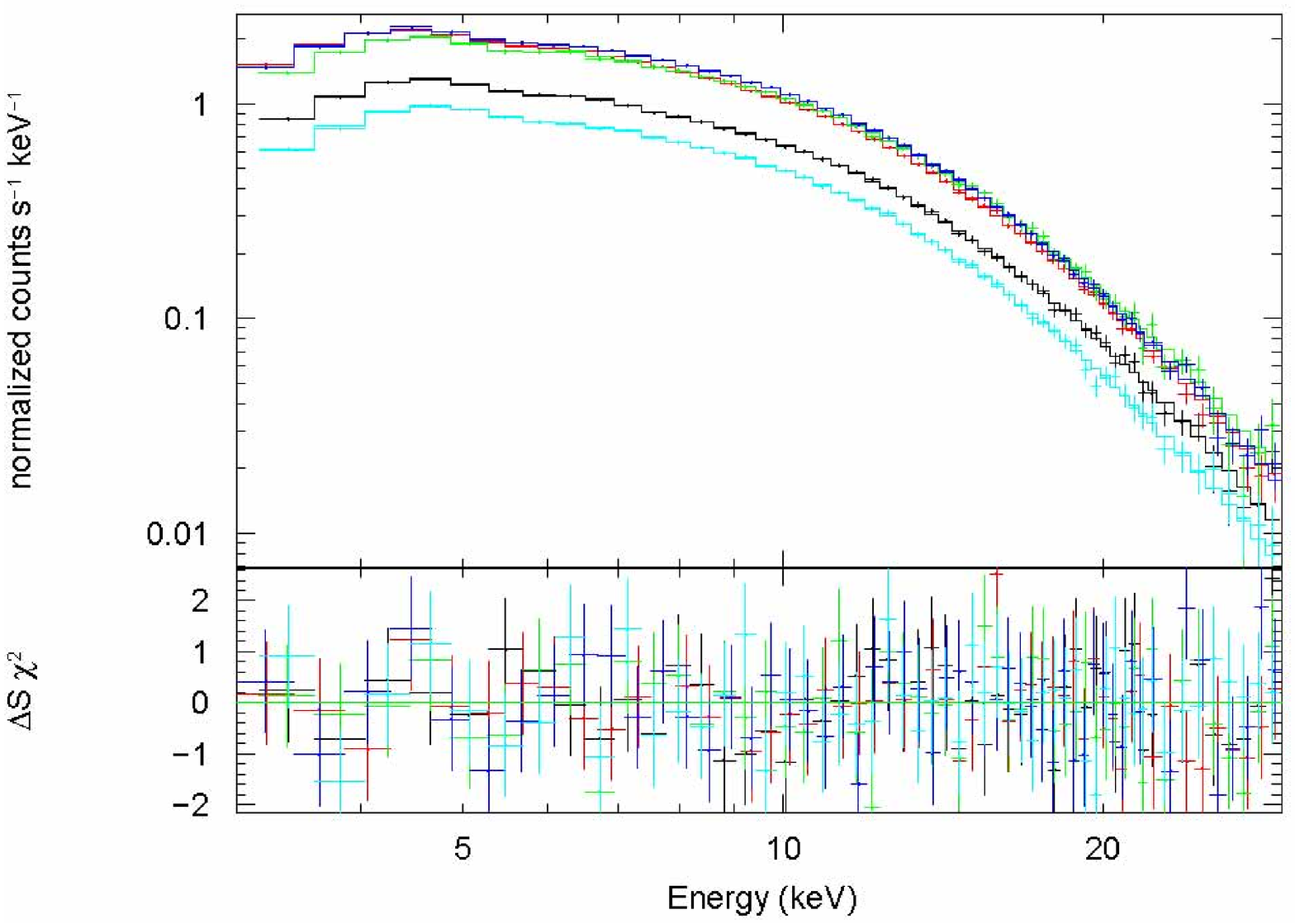}
\caption{PCA 3 -- 30 keV spectra from the orbital phase resolved
spectral analyses, ordered from bottom to top at 5 keV: 0, 0.2,
0.6, 0.4, 0.8.}\label{orbspec}
\end{center}
\end{figure*}

\begin{figure*}
\begin{center}
\includegraphics[width=14cm]{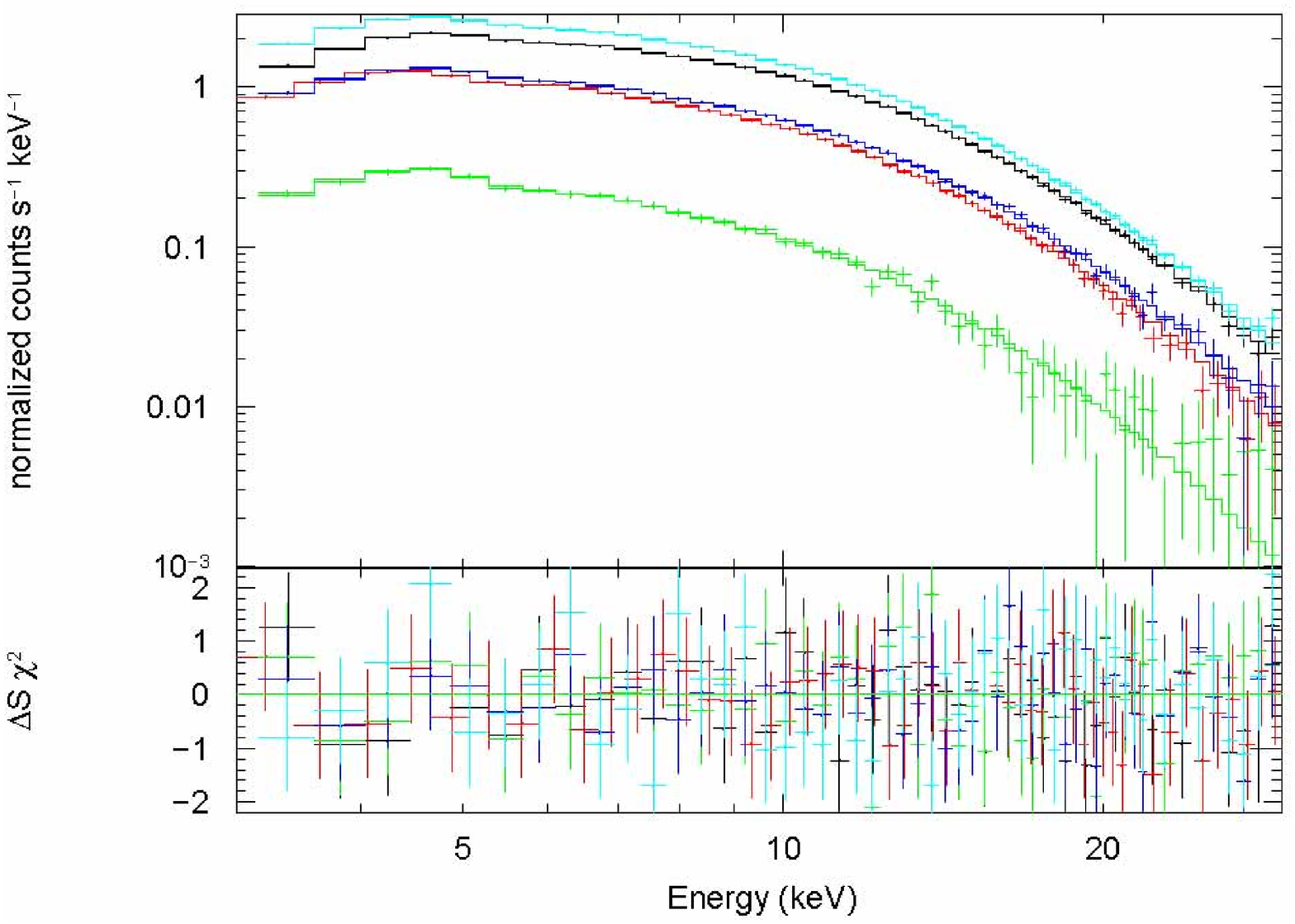}
\caption{PCA 3 -- 30 keV spectra from the super-orbital phase
resolved spectral analyses ordered from bottom to top at 5 keV: 0,
0.8, 0.2, 0.6, 0.4.}\label{sorbspec}
\end{center}
\end{figure*}

\label{lastpage}

\end{document}